\documentclass[reprint,aip,graphicx]{revtex4-1}
\usepackage{chemarr}
\usepackage{graphicx}
\usepackage{graphics}
\usepackage{epstopdf}
\usepackage{xcolor}
\usepackage{float,afterpage}

\draft 

\begin{document}


\title{A general moment expansion method for stochastic kinetic models} 



\author{Angelique Ale}
\thanks{These authors contributed equally.}
\affiliation{Division of Molecular Biosciences, Theoretical Systems Biology Group, Imperial College London, London SW7 2AZ, United Kingdom}
\author{Paul Kirk}
\thanks{These authors contributed equally.}
\affiliation{Division of Molecular Biosciences, Theoretical Systems Biology Group, Imperial College London, London SW7 2AZ, United Kingdom}
\author{Michael P.H. Stumpf}
\email[]{To whom correspondence should be addressed:\newline a.ale@imperial.ac.uk, m.stumpf@imperial.ac.uk}
\affiliation{Division of Molecular Biosciences, Theoretical Systems Biology Group, Imperial College London, London SW7 2AZ, United Kingdom}


\date{\today}

\begin{abstract}
Moment approximation methods are gaining increasing attention for their use in the approximation of the stochastic kinetics of chemical reaction systems. In this paper we derive a general moment expansion method for any type of propensities and which allows expansion up to any number of moments. For some chemical reaction systems, more than two moments are necessary to describe the dynamic properties of the system, which the linear noise approximation (LNA) is unable to provide. Moreover, also for systems for which the mean does not have a strong dependence on higher order moments, moment approximation methods give information about higher order moments of the underlying probability distribution. We demonstrate the method using a dimerisation reaction, Michaelis-Menten kinetics and a model of an oscillating p53 system.  We show that for the dimerisation reaction and Michaelis-Menten enzyme kinetics system higher order moments have limited influence on the estimation of the mean, while for the p53 system, the solution for the mean can require several moments to converge to the average obtained from many stochastic simulations. We also find that agreement between lower order moments does not guarantee that higher moments will agree. Compared to stochastic simulations our approach is numerically highly efficient at capturing the behaviour of stochastic systems in terms of the average and higher moments, and we provide expressions for the computational cost for different system sizes and orders of approximation. {We show how the moment expansion method can be employed to efficiently {quantify parameter sensitivity}.} Finally we investigate the effects of using too few moments on parameter estimation, and provide guidance on how to estimate if the distribution can be accurately approximated using only a few moments.
\end{abstract}

\pacs{82.39.-k, 87.10.Mn, 82.20.Fd, 87.16.A-}

\maketitle 

\section{Introduction}
Cellular behaviour is shaped by molecular process that can be described by systems of chemical reactions between different molecular species. At the macroscopic scale, the dynamics of these processes are frequently described in terms of mean concentrations of species using deterministic mass action kinetics (MAK). The deterministic solution, however, does not always capture the essential kinetics of the chemical system accurately, because it excludes stochastic effects.

The stochastic kinetics of chemical reaction systems are captured by the chemical master equation (CME), a probabilistic description of the system (also known as Kolmogorov's forward equation) \cite{Gardiner2009,Kampen:1992aa}. The CME is a set of differential or difference equations that capture how the probability over the states (e.g. the abundances of the different molecular species) evolves over time. The CME offers an exact description of a system's dynamics but can only be solved analytically for very simple systems. Exact single realisations of the CME, corresponding e.g. to observing processes inside a single cell, can be obtained numerically using for example Gillespie's Stochastic Simulation Algorithm (SSA) \cite{Gillespie1976}. The SSA is a discrete simulation method that proceeds by randomly selecting a reaction that occurs at each subsequent time step, according to the probability of that reaction occurring next. This method is associated with considerable computational cost, that increases dramatically with the size of the model, which can make it infeasible to comprehensively characterize large systems through simulations.

A number of numerical methods have been developed to approximate the solution of the CME for more complex systems, including methods that approximate the CME by describing the probability distribution in terms of its moments {\cite{Gillespie2009,Gomez2007,Lee2009,Singh2010,Singh2006,Ruess2011,Goutsias2007,Barzel2012}}. When only the mean (the first moment) is taken into account, the moment expansion reduces to the MAK description. In the linear noise approximation (LNA), the CME is approximated by taking into account the mean (the first moment), and the variance and covariance (second central moments) of the distribution, whereby the second central moments are decoupled form the mean\cite{Komorowski2009,Komorowski2010,Pahle2012,Wallace2012}. This is valid in the limit of large volumes and molecule numbers, or when the system consists of first-order reactions only. 

For smaller systems and more complex reactions, a number of different approaches have been developed that aim to  {capture} the temporal changes in coupled moments. These expansions can be performed in terms of the moments or the central moments about the mean; the conversion from molecule numbers can be kept implicit in the rate parameters or made more explicit from the outset.
\textcolor{black}{Previous developments focussed on expansion methods for polynomials\cite{Singh2011,Gillespie2009,Sotiropoulos2011}, or expansions up to a limited order. Gillespie et al. developed an expansion method in terms of moments for polynomial functions \cite{Gillespie2009}; this was later generalized to include rational functions as well\cite{Milner2011}. Other alternative methods make the dependence on system size more obvious and include so-called Mass Fluctuation Kinetics (MFK) \cite{Gomez2007}, and the  2MA and 3MA approaches \cite{Ullah2009}.} \textcolor{black}{In the Mass Fluctuation Kinetics approach} the expansion is done in terms of central moments up to third order for first and second order rate equations, and up to second order for general rate equations. In a recent paper \cite{Grima2012} it was shown that expansion up to three moments tends to deliver more accurate results than expansion up to only second order; in particular the variances are improved by going to higher moments. While the MFK, 2MA and 3MA approaches include expressions for the first, second and third central moments, no general formulation exists in these frameworks to generate higher order central moments in an automatic way.
\par
In this paper we derive a general method for expanding the CME in terms of its central moments about the mean, which does not make extraneous assumptions about the form of the reactions, and that can be evaluated up to any number of moments;  in our exposition we follow the notation used in \citet{Gillespie2009}. {The new  method described here non-trivially generalizes the work of Gomez et al.\cite{Gomez2007}: in particular we are able to generate arbitrary order higher central moments automatically and in a computationally efficient manner; combining computer algebra systems with numerical simulation engines does allow us to tackle problems that stymy e.g. the linear noise approximation or conventional (3rd order) MFK approaches. The interplay between noise and non-linear dynamics can give rise to very complicated behaviour, and only a handful of systems have been considered. The moments of the distribution described by the CME reflect this intricate relationship between stochasticty and non-linearity and we spend some time discussing this for a typical non-linear biomolecular system. Thus while going to arbitrary moments is straightforward in our framework, we focus our discussion on the lower (up to sixth order) moments.}
\par
{Our expansion is based on two successive Taylor expansions that allow us to express the CME of the moment generating function in computationally convenient form}; and we truncate the system by setting higher order terms in the Taylor expansions to zero. The outlined method {could}  fit into a general framework for parameter inference based on maximum entropy distributions derived from the calculated moments.  {We furthermore demonstrate that parameter sensitivity analyses may be performed naturally and efficiently in this framework:  we can consider the rate of change of the moments with respect to the parameters, which also allows us to study the factors underlying cell-to-cell variability.}  
  We illustrate the {general} method using three molecular reaction systems: a simple dimerisation reaction, which allows for a detailed investigation;  Michaelis-Menten enzyme kinetics; and the p53 system, an oscillating tumour suppressor system \cite{Batchelor:2009hk}. 

\section{Moment expansion method}
We consider a system with $N$ different molecular species $(X_1,...X_N)$ that are involved in $L$ chemical reactions
with reaction rates $k_l$,
\begin{eqnarray}
\underbar{s}_1X_1+...+\underbar{s}_NX_N \xrightarrow{k_l}  \bar{s}_1X_1+...+\bar{s}_NX_N,
\end{eqnarray}
with $\underbar{s}_i$ and $\bar{s}_i$ the number of molecules of type $X_i$ before and after the reaction, respectively.
The time evolution of the system's state  is described by the chemical master equation (CME),
\begin{eqnarray}
\frac{dP(\textbf{x})}{dt}=\sum_{l}P(\textbf{x}-\textbf{s}_l)a_l(\textbf{x}-\textbf{s}_l)-P(\textbf{x})a_l(\textbf{x}),
\end{eqnarray}
in which $a_l(\textbf{x})$ are the propensity functions, with $a_l(\textbf{x})dt$ defined as the probability of reaction $l$ occurring in an infinitesimal time interval $dt$ when the number of molecules in the system is $\textbf{x}$ , $P(\textbf{x})$ the probability that the system contains $\textbf{x}$ molecules and $\textbf{s}_{l}=\bar{\mathbf{s}}_{l}-\underbar{\textbf{s}}_{l}$.
\par
We start the derivation of the moment expansion method by deriving a moment generating function from the CME. In general, the moment generating function of a random variable \textbf{x} is defined as \cite{Gardiner2009}
\begin{eqnarray}
m(\theta,\mathbf{x})=\sum_{\textbf{x}}e^{\theta \textbf{x}}P(\textbf{x}).
\end{eqnarray}
The moments, $\left<\textbf{x}^\textbf{n}\right>$, with $\textbf{x}^\textbf{n}=x_1^{n_1}...x_d^{n_d}$, of the probability distribution can be found by taking the $n$-th order derivatives of the moment generating function with respect to $\theta$. The first moment is, of course, equal to the mean, $\mu=\left<x\right>$. The variance, skewness (related to asymmetry of the distribution) and kurtosis (related to the chance of outliers) can be derived from the central moments about the mean, $\left<(\textbf{x}-\mu)^\textbf{n}\right>$.
\par
Using the CME we can write the time dependent moment generating function \cite{Azunre2007}
\begin{eqnarray}
\frac{dm}{dt}=\sum_{l}\left[(e^{\theta \textbf{s}_l}-1)\sum_{\textbf{x}}e^{\theta \textbf{x}}P(\textbf{x})a_l(\textbf{x})\right]
\label{mgf}
\end{eqnarray}
The time evolution of the mean concentration $\mu_i$ of species $X_i$ can be obtained by
taking the first derivative of Eq. \ref{mgf} with respect to $\theta_i$ and subsequently setting $\theta$ to zero,
\begin{eqnarray}
\frac{d\mu_{i}}{dt}=\left.\frac{d}{d\theta_i}\frac{dm}{dt}\right|_{\theta=0}= \sum_{l}s_l\left<a_l(\mathbf{x})\right>
\end{eqnarray}
This expression can be evaluated by expanding $a_l(\textbf{x})$ in a Taylor expansion about the mean, 
\begin{eqnarray}
\frac{d\mu_i}{dt}=S\left[\sum_{l}\left.\sum_{n_1=0}^\infty...\sum_{n_d=0}^\infty\frac{1}{\mathbf{n!}}\frac{\mathbf{\partial^{n}}a_l(\textbf{x})}{\partial\mathbf{x^{n}}}\right|_{x=\mu}\mathbf{M_{x^{n}}}\right],
\label{eq6}
\end{eqnarray}
where $S$ is the stoichiometry matrix, 
\begin{eqnarray}
\mathbf{\partial^{n}}=\partial^{n_1+..+n_d}\nonumber\\
\mathbf{n!}=n_1!...n_d!\nonumber\\
\mathbf{M_{x^{n}}}=M_{x_1^{n_1},...,x_d^{n_d}}\nonumber\\
\mathbf{\partial x^n}=\partial x_1^{n_1}...\partial x_d^{n_d},\nonumber
\label{mmgf}
\end{eqnarray}
and we have substituted the central moments around the mean for the expected values of the expansion terms
\begin{eqnarray}
M_{x_1^{n_1},...,x_d^{n_d}}=\left<(x_1-\mu_1)^{n_1}...(x_n-\mu_n)^{n_2}\right>.\nonumber
\label{mmgf}
\end{eqnarray}
The first central moment $\left<\mathbf{x}-\mu\right>$ is zero. Higher order central moments can be derived from the moments, for example the covariance between $x_1$ and $x_2$ can be written as
\begin{eqnarray}
\sigma_{x_1x_2}^2=\left<(x_1-\mu_{1})(x_2-\mu_{2})\right>=\nonumber\\
\left<x_1x_2\right>+\mu_1\mu_2-\left<x_1\right>\left<\mu_2\right>-\left<x_2\right>\left<\mu_1\right>.
\label{cmgf2}
\end{eqnarray}
In general the relation between the central moments, $\mathbf{M_{x^{n}}}$, and the moments, $\mathbf{\mu^n}$, can be formulated as
\begin{eqnarray}
\mathbf{M_{x^{n}}}=\sum_{k_1=0}^{n_1}...\sum_{k_d=0}^{n_d}\binom{\mathbf{n}}{\mathbf{k}}(-1)^{\mathbf{(n-k)}}\underbrace{\mathbf{\mu^{(n-k)}}}_{\alpha}\underbrace{\left<\mathbf{x^{k}}\right>}_{\beta},
\label{eq7}
\end{eqnarray}
where
\begin{eqnarray}
(-1)^{\mathbf{(n-k)}}=(-1)^{(n_1-k_1)}...(-1)^{(n_d-k_d)}\nonumber\\
\mathbf{\mu^{(n-k)}}=\mu_1^{(n_1-k_1)}...\mu_d^{(n_d-k_d)}\nonumber\\
\mathbf{x^k}=x_1^{k_1}...x_d^{k_d}\nonumber\\
\binom{\mathbf{n}}{\mathbf{k}}=\binom{n_1}{k_1}...\binom{n_d}{k_d}.\nonumber
\label{eq8}
\end{eqnarray}
We obtain the time evolution equations of the central moments by taking the time derivative of Eq. \ref{eq7},
\begin{eqnarray}
\frac{d\mathbf{M_{x^{n}}}}{dt}=\sum_{k_1=0}^{n_1}...\sum_{k_d=0}^{n_d}\binom{\mathbf{n}}{\mathbf{k}}(-1)^\mathbf{{(n-k)}}\left[\alpha\frac{d\beta}{dt}+\beta\frac{d\alpha}{dt}\right].
\label{cmgf}
\end{eqnarray}
The term $\alpha$ makes the time evolution of the central moments a function of Eq. \ref{eq6}, 
\begin{eqnarray}
\frac{d\alpha}{dt}=\sum_{i=1}^N(n_i-k_i)\mu_i^{-1}\alpha\frac{d\mu_i}{dt}
\end{eqnarray}
The term $\beta$ gives rise to mixed moments, and the derivative of $\beta$ with respect to time yields the time evolution equations for the mixed moments. Therefore we also need to include the time derivatives of the mixed moments in our system of equations. We obtain the time derivative of the term $\left<\mathbf{x^{k}}\right>$ by taking higher order derivatives of the moment generating function Eq. \ref{mgf}, resulting in 
\begin{eqnarray}
\frac{d\beta}{dt}=\sum_{e_1=0}^{k_1}...\sum_{e_d=0}^{k_d}\mathbf{s^{e}}\binom{\mathbf{k}}{\mathbf{e}}\underbrace{\left<\mathbf{x^{(k-e)}}a(x) \right> }_{\left<F\right>},
\end{eqnarray}
where
\begin{eqnarray}
\mathbf{s^e}=s_{1}^{e_1}...s_{d}^{e_d}\nonumber\\
\mathbf{x^{(k-e)}}=x_1^{k_1-e_1}...x_d^{k_d-e_d}.\nonumber
\label{mmgf}
\end{eqnarray}
By expanding the individual terms of the resulting expressions in a second Taylor expansion, the time evolution of the mixed moments becomes a function of the central moments; the time evolution of the central moments remains a function of the central moments alone,
\begin{eqnarray}
\left<F\right>=\left.\sum_{n_1=0}^\infty...\sum_{n_d=0}^\infty\frac{1}{\mathbf{n!}}\frac{\mathbf{\partial^{n}}F(x)}{\mathbf{\partial x^{n}}}\right|_{x=\mu}{\mathbf{M_{x^{n}}}}.
\end{eqnarray}
\par
When the model under investigation is non-linear, each central moment will depend on a higher order central moment, which may itself also depend on higher order moments; hence the number of equations we would need to include is in principle infinite.
To overcome this, we can obtain a closed set of equations by evaluating the time evolution equations for $\nu$ moments and truncating the Taylor series after the $\nu$th order, thereby setting all higher order {central moments equal to zero. 
By truncating the Taylor expansion (i.e. setting terms of the Taylor expansion corresponding to $\sum n_i>\nu$ to 0), the equations are only dependent on the central moments up to the selected order $\nu$.}
Alternatively, the set of equations could be closed using moment closure techniques based on common expressions for well known probability distributions \cite{Milner2011}. In this paper we will use truncation as well as closure based on a Gaussian distribution.                                                                                                        
\vspace{-\baselineskip}
\section{Results}
We illustrate the approach in a range of applications that serve to highlight both the basic properties of the moment expansion method as well as the advantages this method offers in situations where other methods \cite{Ito2010,Komorowski2012,Wallace2012} tend to fail.
\vspace{-\baselineskip}
\subsection{Dimerisation}
We first illustrate the moment approximation method using a simple dimerisation reaction \cite{Wilkinson2012}. The system describes the reversible formation and disintegration of a dimeric molecule,
\begin{eqnarray}
X_1+X_1 \xrightleftharpoons[k_2]{k_1} X_2.
\end{eqnarray}
The system can be written in terms of two propensities
\begin{eqnarray}
\begin{array}{cc}
a_1: k_1x_1(x_1-1);&\hspace{7mm} a_2: k_2x_2,
\end{array}
\end{eqnarray}
and the stoichiometry matrix
\begin{eqnarray}
S=\left[\begin{array}{cc}
-2 & 2\\
1 & -1\end{array} \right],
\end{eqnarray}
where the columns correspond to reactions and the rows to variables.
\par
\begin{figure}
 \includegraphics[width=0.47\textwidth]{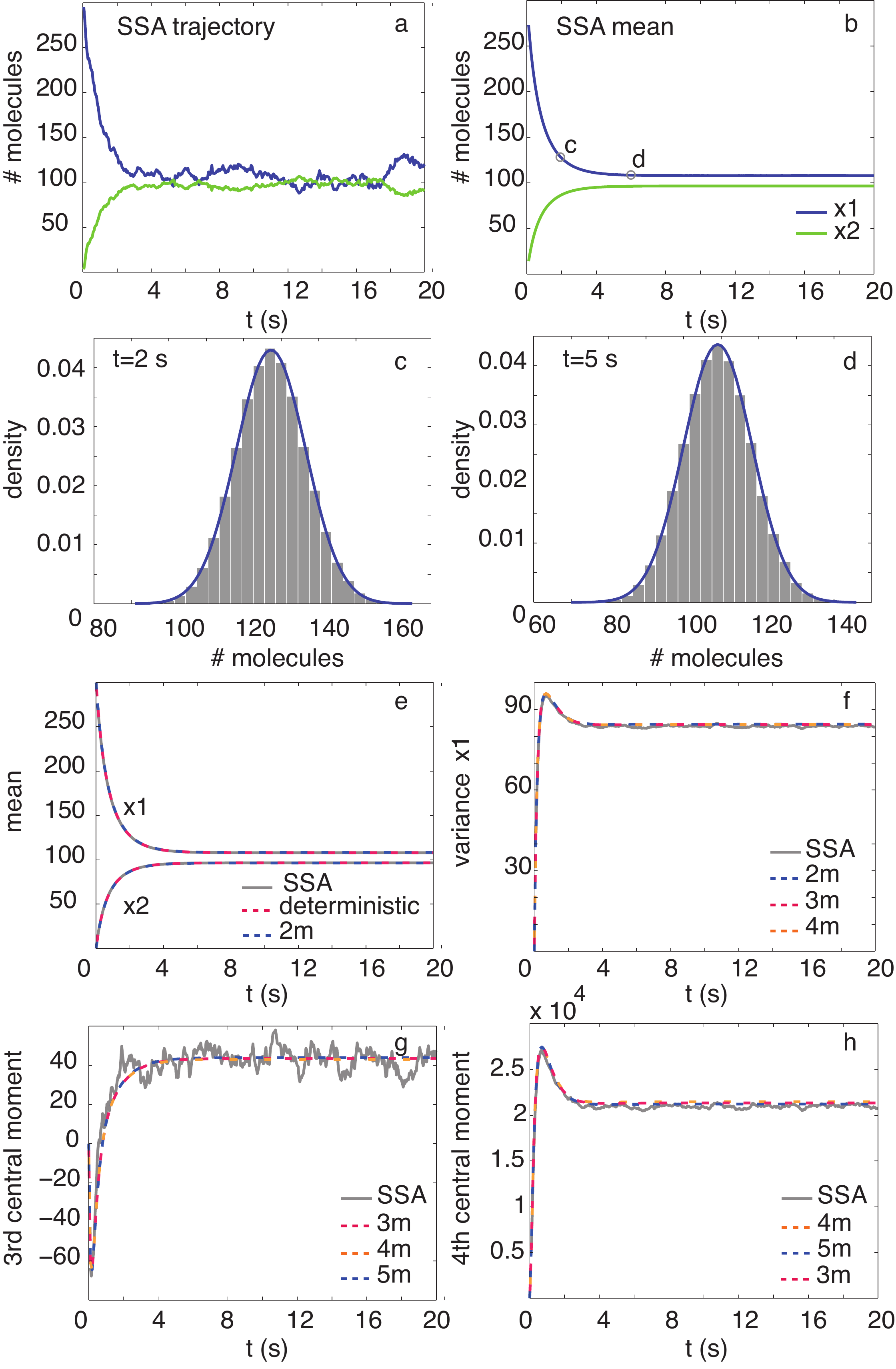}
 \caption{\label{}Study of dimerization system, initial values $\mathbf{x}=[301,0]$, parameters $\mathbf{k}=[1.66\cdot10^{-3}, 0.2]$. a) Single SSA realisation. b) Average of 100,000 SSA simulations. c-d) Histogram of SSA runs (grey bars) and probability density of normal distribution (blue line) calculated from mean and variance of SSA runs corresponding to points c and d in figure (b). e) Mean for both variables, calculated using SSA, the moment approximation using 1 moment (deterministic) and two central moments (2m). f) Variance of $x_1$ calculated using SSA, two central moments (2m), three central moments with Gaussian closure (3m) and four central moments (4m). g) Third central moment calculated using SSA and moment approximation method, fourth central moment calculated using SSA and moment approximation method.  }
 \label{dimer1}
 \end{figure}

The exact kinetics of the system can be straightforwardly simulated using the Stochastic Simulation Algorithm (SSA) \cite{Gillespie1976}. One realisation of the SSA is equivalent to observing the kinetics of the system inside a single cell (Figure \ref{dimer1}a), whereas the average over many realisations mimics the observation of the average kinetics for a large number of cells (Figure \ref{dimer1}b, n=100,000). The system reaches a stationary state after about $4$ seconds.
In Fig. \ref{dimer1}c-d we plotted histograms of the SSA at two time points that correspond to different regimes in the trajectory:  transient state with decreasing molecule number (Fig. \ref{dimer1}c) and stationary state (Fig. \ref{dimer1}d). We calculated the normal probability density function from the mean and variance of the SSA distribution at those two time points (blue line), ignoring higher order moments. The normal probability density functions fit the histograms very well, which indicates that the distribution of molecule numbers is approximately normal over the time course, and higher order moments would have limited influence on describing the kinetics of the mean for this system.
\par
Figure \ref{dimer1}d shows the mean molecule numbers calculated with the moment approximation method compared to the SSA results. The results for the deterministic (including only the mean) as well as the two moment approximation are approximately equal to the means calculated with the SSA. We compare the higher order central moments calculated with the general moment approximation method described above with the results from the SSA simulations in Figures \ref{dimer1}f-h. In the 3--moment approximation we closed the equations using the Gaussian probability distribution (setting the fourth cumulant equal to zero \cite{Grima2012}), while in the other approximations we truncated higher order moments. The approximated moments are close to the exact moments calculated from the SSA, which is also clear from the errors displayed in Table \ref{table1}, calculated as $\epsilon=(1/N)\sum_{n} \sqrt{((M_{SSA}-M_{ma})/M_{SSA})^2}$ with $M_{SSA}$ the moment or central moment calculated based on the SSA, $M_{ma}$ the corresponding value calculated with the moment approximation, and $N$ the number of time points taken into account. The larger error for the mean when using the deterministic approach is due to small differences in the first part of the trajectory, the decreasing part, where the contributions of the higher order central moments are largest. {The larger errors calculated for the third central moment are due to the fluctuations that are still present in the third central moment trajectory calculated based on 100,000 SSA runs. Increasing the number of SSA runs would reduce this effect.}

\begin{table}
 \caption{\label{table1}Error between mean, second and third central moment calculated with SSA and approximation methods for the dimerization system.}
 
\vspace{5mm}
\begin{tabular}{ l |c | c | c | c | c | c  c }
\hline \hline
 $\epsilon [\%]$& deterministic & 2m & 3m & 4m & 5m & 6m\\
 \hline 
$\mu_1$ & 0.300 & 0.0545 & 0.0546& 0.0546 & 0.0546&0.0546 \\
 $M_{x_1^2}$ & & 0.961 & 0.803& 0.804 &0.801 & 0.803\\
 $M_{x_1^3}$ & & & 18.3 &  18.1&18.2&18.2\\
 $M_{x_1^4}$ & & & & 2.39 &1.33 &1.83\\
 \hline \hline
\end{tabular}
\end{table}

\subsection{Michaelis-Menten enzyme kinetics}
We next look at Michaelis-Menten enzyme kinetics, where an enzyme, $E$, and substrate, $S$, first bind to form a complex,  $SE$; following this, the complex can dissociate into the original components $S$ and $E$, or $S$ can be converted into the product, $P$,
\begin{eqnarray}
S+E \xrightarrow{k_1} SE\nonumber\\
SE \xrightarrow{k_2} S+E\\
SE \xrightarrow{k_3} P+E\nonumber
\end{eqnarray}
The system is often reduced to a system of two variables ($S$ and $P$) \cite{Wilkinson2012}, with three reaction propensities
\begin{eqnarray}
\begin{array}{c}
a_1: k_1S[E(0)-S(0)+S+P];\nonumber\\
a_2: k_2[S(0)-(S+P)];\nonumber\\
a_3: k_3[S(0)-(S+P)]
\end{array}
\end{eqnarray}
and the stoichiometry matrix
\begin{eqnarray}
S=\left[\begin{array}{ccc}
-1 & 1 & 0\\
0 & 0 & 1\end{array} \right].
\end{eqnarray}

\begin{figure}
\includegraphics[width=0.48\textwidth]{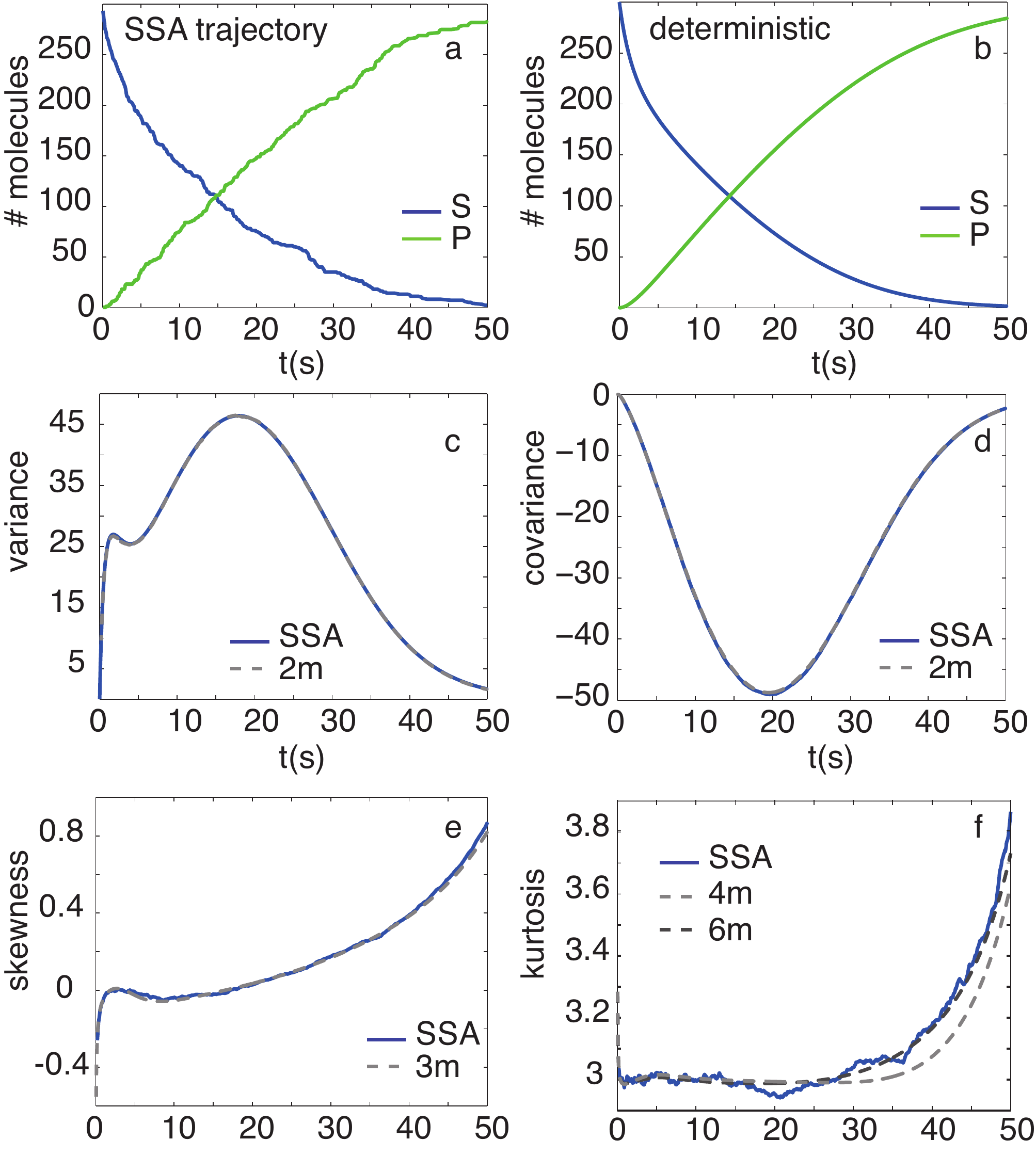}
\caption{\label{}Study of Michaelis-Menten kinetics, with parameters $\textbf{k}=[1.66\cdot10^{-3},1\cdot10^{-4},0.01]$, and initial conditions, $S(0)=301$, $P(0)=0$, and  $E(0)=120$. (a) Single SSA realisation. Trajectories calculated using (b) moment approximation including only the mean (deterministic). (c-d) Variance of S and covariance between S and P calculated using SSA and approximation using 2 moments. (e) Skewness of S calculated using SSA and approximation with 3 moments, (f) Kurtosis calculated using SSA and approximation up to 4 and 6 moments. }
\label{mmfigure1}
\end{figure}
One trajectory calculated with Gillespie's Stochastic Simulation Algorithm is shown in Figure \ref{mmfigure1}a, and the mean calculated by solving the ODE system using the deterministic representation of the system is displayed in Figure \ref{mmfigure1}b. For this system the deterministic representation is very close to the stochastic solution. The variance of the substrate and the covariance between the substrate and the product (Figure \ref{mmfigure1}c-d) calculated based on 100,000 SSA runs can be closely approximated using the moment approximation method expanded up to two moments. Figure \ref{mmfigure1}e shows the skewness of the distribution over time, calculated as
\begin{eqnarray}
\gamma=\frac{\left<(x_1-\mu_1)^3\right>}{\sigma_{11}^3}=\frac{\left<(x_1-\mu_1)^3\right>}{\left<(x_1-\mu_1)^2\right>^{3/2}}.
\end{eqnarray}
For a normal distribution the skewness is zero. The skewness is approximated well using the moment approximation method up to three moments. The kurtosis, given by 
\begin{eqnarray}
\gamma_2=\frac{\left<(x_1-\mu_1)^4\right>}{\sigma_{11}^4}=\frac{\left<(x_1-\mu_1)^4\right>}{\left<(x_1-\mu_1)^2\right>^{4/2}},
\end{eqnarray}
indicates the thickness of the tails of the probability distribution, relating to the probability of outliers. For a normal distribution, the kurtosis is equal to 3. When four moments are used to approximate the system, the approximation of the kurtosis that we obtain from the SSA is not as close as when also higher moments, here six moments, are included in the calculation. This illustrates that agreement between lower-order moments does not guarantee that higher-order moments will also agree. This problem is likely exacerbated for more complex models, e.g. those exhibiting non-linear dynamics.
 \afterpage{\begin{figure}[H]
 \includegraphics[width=0.475\textwidth]{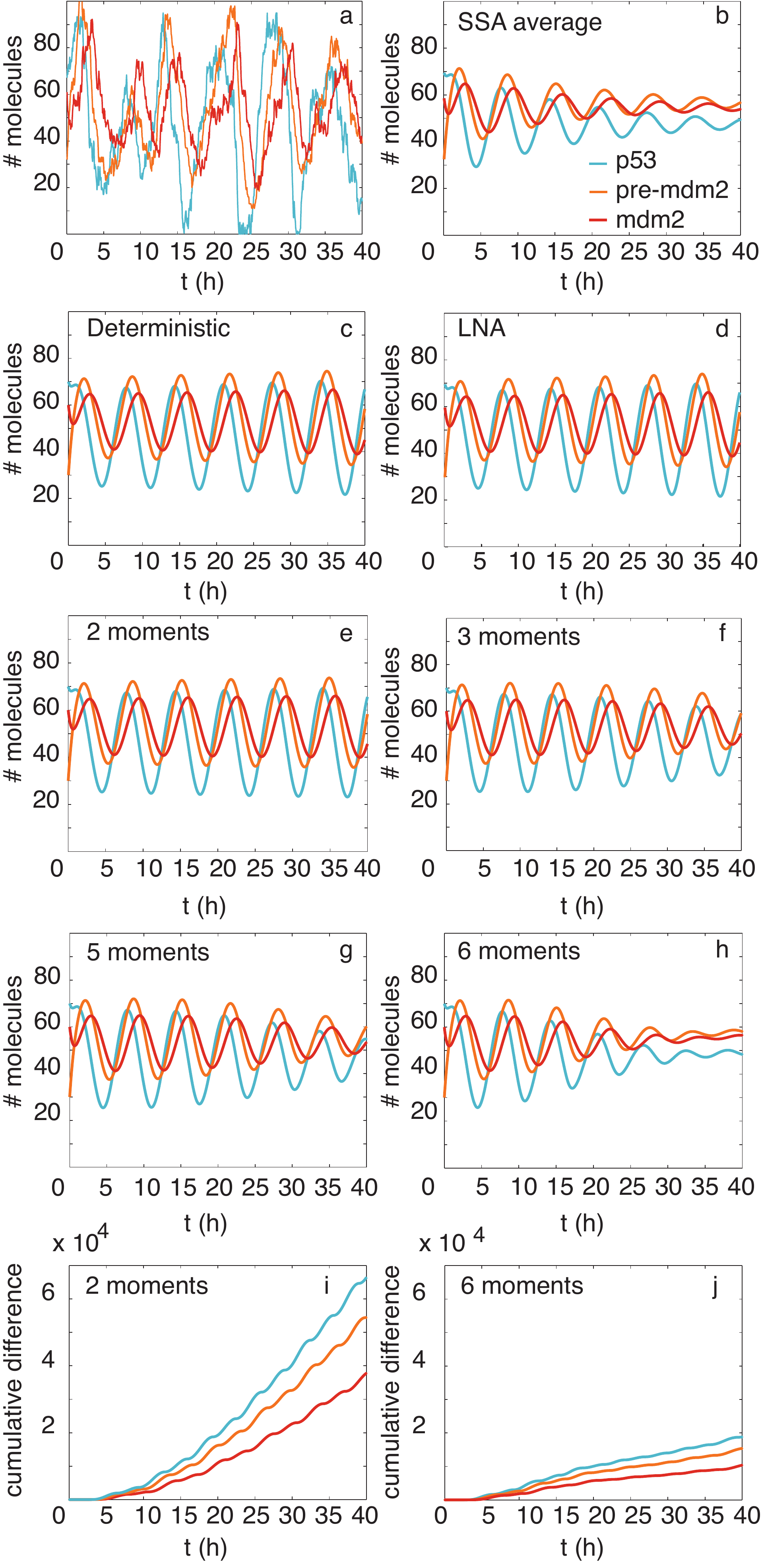}
 \caption{Study of p53 model, parameter set $q_1=[90,0.002,1.7,1.1,0.93,0.96,0.01]$. (a) Single SSA realisation. (b) Average of 100.000 SSA simulations. Trajectories calculated using (c) moment approximation including only the mean (deterministic), (d) Linear Noise Approximation, (e) mean and the variance, (f) up to three central moments, (g) five central moments and (h) six central moments. (i-j) Cumulative difference between mean trajectory calculated with SSA and trajectories calculated using (i) 2 moments and (j) 6 moments.}
 \label{p53figure1}
 \end{figure}}
 \subsection{P53 system}
Finally, we investigate the oscillatory p53 system\cite{Geva2006}, which consists of three proteins connected through a nonlinear feedback loop: p53, precursor of Mdm2 and Mdm2. The system can be written in terms of six propensities, 
\begin{eqnarray}
\begin{array}{cc}\vspace{2mm}
a_1: k_1;&\hspace{7mm} a_2: k_2x\\\vspace{2mm}
a_3: k_{3}\frac{xy}{x+k_{7}};&\hspace{7mm} a_4: k_4x\\
a_5: k_5y_0; &\hspace{7mm} a_6:k_6y,
\label{p53system}
\end{array}
\end{eqnarray}
and the stoichiometry matrix
\begin{eqnarray}
S=\left[ \begin{array}{cccccc}
1 & -1 & -1 & 0& 0& 0\\
0 & 0 & 0 & 1& -1 & 0\\
0 & 0 & 0 & 0 & 1 & -1\end{array} \right],
\end{eqnarray}
where  $x$ is the concentration of p53, $y_0$ the concentration of precursor of Mdm2, $y$ is the concentration of Mdm2,  $k_1$ is the p53 production rate, $k_2$ is the Mdm2-independent p53 degradation rate, $k_{3}$ the saturating p53 degradation rate, $k_{7}$ is the p53 threshold for degradation by Mdm2, $k_{4}$ is the p53-dependent Mdm2 production rate, $k_5$ is the Mdm2 maturation rate and $k_6$ is the Mdm2 degradation rate.
\par
Figure \ref{p53figure1}a shows an SSA simulation of the model for parameters $q_1=$[90,0.002,1.7,1.1,0.93,0.96,0.01] and initial values $\textbf{x}(0)=[70, 30, 60]$, which exhibits oscillating behaviour. Because the oscillations for different realisations of the SSA (corresponding to different cells) are stochastically out of phase, the average over 100,000 stochastic simulations shows a damped oscillation, reflecting the presence of a negative feedback loop. Figures \ref{p53figure1}c-h show the trajectories of the mean calculated with the moment approximation method. In the deterministic approximation, the oscillations are not damped but expanding, which would indicate a positive instead of negative feedback loop. The LNA and the 2 moment approximation show the same effect. The mean calculated with the LNA is always equal to the mean calculated with the deterministic approximation because the mean is not coupled to the variance. When 3 moments are included, the system shows damped oscillations, although not as damped as the SSA trajectories. By including more moments the trajectories converge to the SSA trajectories. When 6 moments are taken into account, the trajectories calculated with the moment approximation show a similar behaviour to the trajectories calculated with the SSA. This is confirmed by the cumulative difference between the SSA trajectories and the trajectories calculated with the moment approximation shown in Figures \ref{p53figure1}i-j, which show a clear decrease in cumulative error for all variables when 6 central moments compared to 2 central moments are included in the approximation. Including more moments would improve the estimation further.
\par
\begin{figure}[t]
 \includegraphics[width=0.48\textwidth]{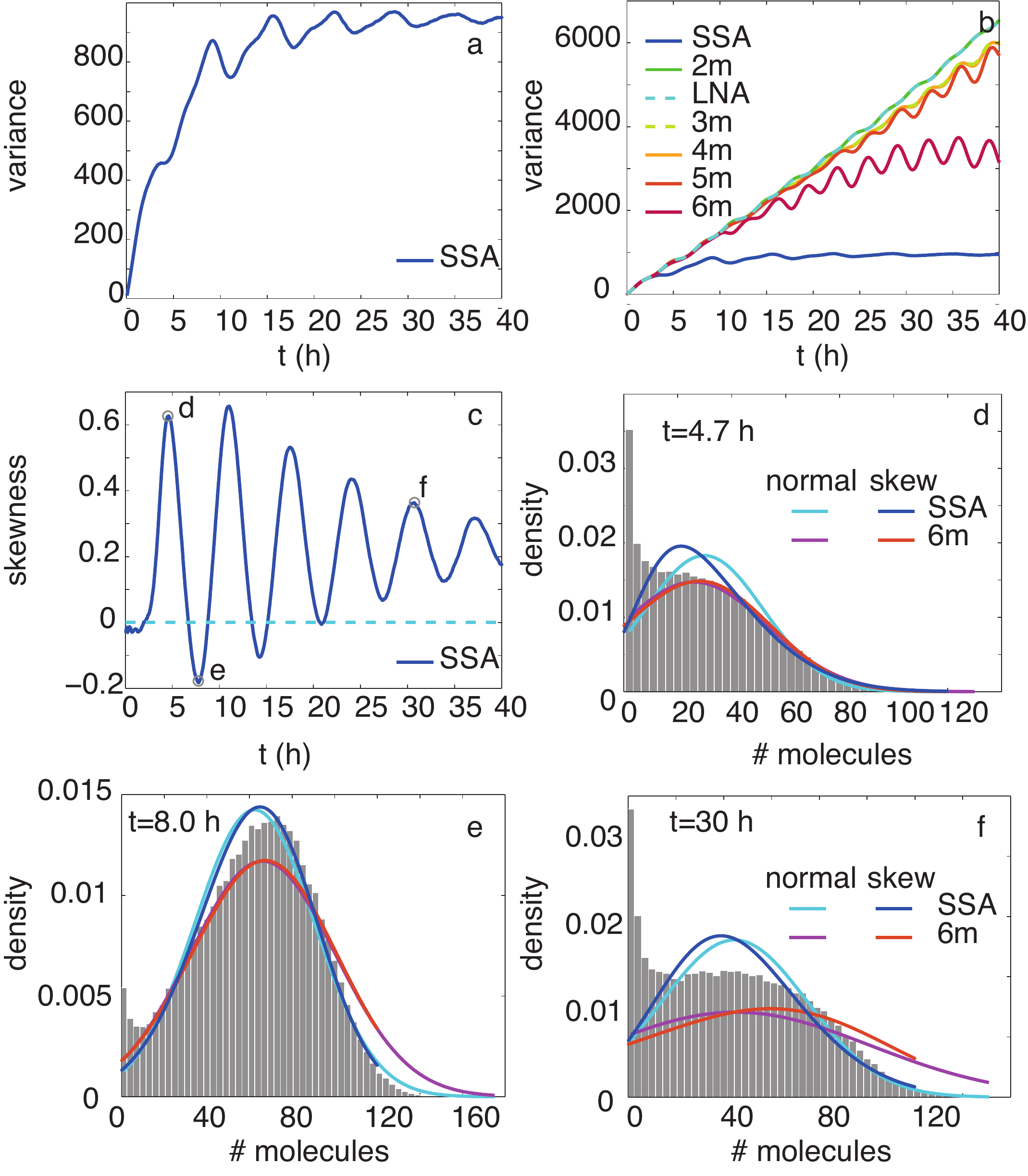}
 \caption{Analysis of distribution for p53 model. (a) Variance calculated based on SSA runs. (b) Variance calculated with SSA, LNA and moment approximation method. (c) Skewness calculated based on SSA runs (blue line) and skewness for normal distribution (cyan dashed line). (d-f) Histograms calculated based on SSA for points d, e and f in figure (c), and probability densitiy function of normal distribution calculated using mean and variance based on SSA (cyan line).}
 \label{p53figure2}
 \end{figure}

We analyze the distribution of the p53 model over the time course by looking at the central moments (Figure \ref{p53figure2}). The variance for the SSA is first increasing, then after about 20 hours it levels off. In Figure \ref{p53figure2}b we compare the variance calculated based on the SSA with the LNA and moment approximation method. When up to five central moments are included, the variance keeps increasing and does not level off. Only for the case of 6 moments does the variance reach a stable value, and even then the value is about three times higher than that predicted by the SSA simulations. Figure \ref{p53figure2}c shows the comparison of the skewness of the SSA distribution to the skewness of a normal distribution ($\gamma=0$). For three time points where the skewness is relatively large (indicated by d, e, f) we display the histogram of the 100,000 SSA realisations together with the probability density of the normal distribution (cyan line) calculated based on the mean and variance of the SSA and the 6 moment approximation. Additionally, we plot the skew-normal distribution calculated using the mean, variance and third central moment; this is defined by the probability density function
\begin{eqnarray}
f(x)=\frac{2}{\omega}\phi\left(\frac{x-\xi}{\omega}\right)\psi\left(\alpha\left(\frac{x-\xi}{\omega}\right)\right), 
\end{eqnarray}
with $\xi$ a location parameter and $\omega$ a scale parameter. The parameter $\delta$ can be calculated from the estimated skewness using the relation
\begin{eqnarray}
|\delta|=\sqrt{\frac{\pi}{2}\frac{|\hat{\gamma}|^{2/3}}{|\hat{\gamma}|^{2/3}+((4-\pi)/2)^{2/3}}}.
\end{eqnarray}
The parameter $\alpha$ can be calculated from $\delta$ with $\delta=\alpha/(\sqrt{1-\delta^2})$, and $\omega$ can be calculated from the variance using $\sigma^2=\omega^2(1-2\delta^2/\pi)$. These plots confirm what we saw above, that using only the mean and variance does not capture the full distribution in this case, and also including the skewness is not enough. {When comparing the skewness of the p53 distribution with the skewness of the Michaelis-Menten enzyme kinetics system (Figure \ref{mmfigure1}e), we find that the maximum value of the skewness for both systems is approximately equal, and in both systems the skewness does not have a large effect on the mean.} 

 \begin{figure}[t]
\includegraphics[width=0.5\textwidth]{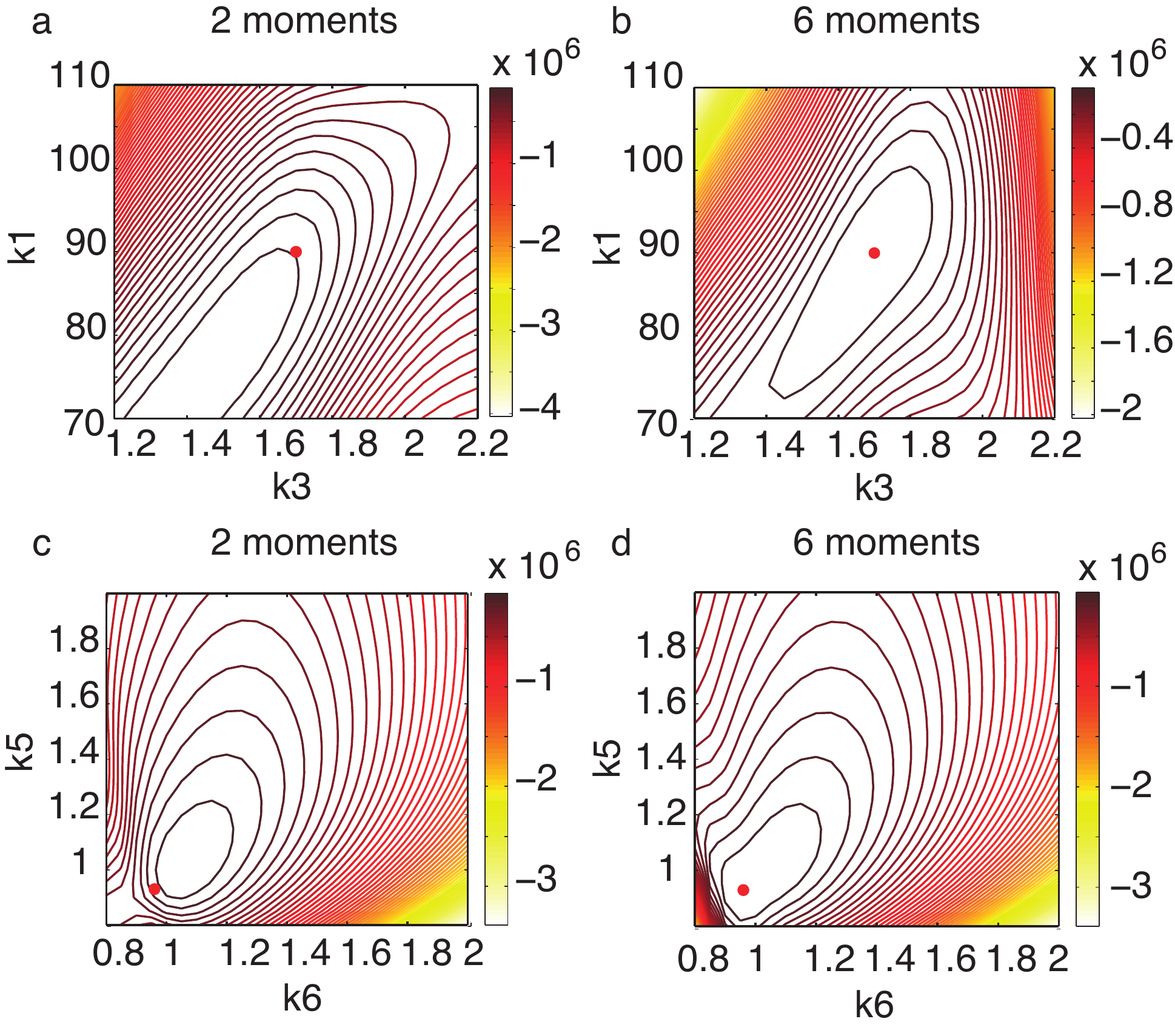}
\caption{\label{}\textcolor{black}{Contour plots  for the p53 system of the distance $d$ between SSA trajectories and trajectories calculated with the moment expansion method. a-b) Contour for varying $k_1$ and $k_3$, calculated using expansion up to 2 and 6 moments respectively. c-d) Contour for varying $k_5$ and $k_6$, calculated using expansion up to 2 and 6 moments respectively. Red dots indicate parameter values used for SSA simulations.}}
\label{contour1}
\end{figure}
\textcolor{black}{We evaluate the influence of ignoring higher order moments on parameter estimation with this system by looking at 2-dimensional contour diagrams of the distance $d$ {(which is closely related to the Gaussian likelihood \citep{Kirk2008})} calculated via }
\textcolor{black}{
\begin{eqnarray}
d=-0.5{\sum_{t=1}^T\sum_{i=1}^3(x_i-\mu_i) ^2}
\end{eqnarray}}
\textcolor{black}{where $t$ are the time points of the trajectories and $T$ the total number of time points included in the calculation ($T=800$), $x_i$ are the mean values of the SSA trajectories for the three variables in the system at timepoints $t$, and $\mu_i$ are the mean values of the trajectories calculated with the moment expansion method. Figures \ref{contour1}a-b show contour plots of the distances between the SSA trajectories and moment expansion trajectories calculated for varying values of parameters  $k_1$ and $k_3$ in the p53 system (Eq. \ref{p53system}), Figures \ref{contour1}c-d show contour plots of the distances calculated for varying values of parameters  $k_5$ and $k_6$. The parameters used for the SSA simulations ($k_1=90$, $k_3=1.7$, $k_5=0.93$, $k_6=0.96$) are indicated with a red dot. From the contour plots corresponding to expansion up to  two central moments (Figures \ref{contour1}a,c) we can see that the minimum distance is obtained for parameters values that are not equal to the actual parameters used for the SSA simulations. This indicates that when expansion up to two moments is used for parameter inference of the p53 system, the wrong parameters will be found. However, when six central moments are included (Figures \ref{contour1}b,d), the minimum distance does correspond to the parameters used for the SSA calculations.}

\subsection{Parameter Sensitivity Estimation}

Assessing parameter sensitivity is a key concern when fitting any parametric model \citep{Saltelli2004,Erguler2011}.  Such analyses allow us to quantify how rapidly the outputs of our model change as we vary its parameters, which can provide insights into the robustness of the model and the relative influence that each parameter has upon the model's behaviour.  However, sensitivity analyses of stochastic models can be difficult and/or computationally costly \citep{Gunawan2005,Plyasunov2007}, and often involve simulating many times in order to obtain Monte Carlo estimates of sensitivity coefficients.  The development of efficient methods for stochastic sensitivity analyses has been the focus of much recent research \citep{Plyasunov2007,Komorowski2012b,Sheppard2012}.    

In the context of our proposed approach, a natural and straightforward way to assess parameter sensitivity is to consider the rate at which the moments vary with the parameters.  This motivates the calculation of simple sensitivity coefficients \citep{Varma1999,Saltelli2004} of the form $s_{ij}(t) = \frac{\partial m_i(t, \boldsymbol{\theta})}{\partial \theta_j}$, where $m_i$ is the (estimated) $i$-th moment and $\theta_j$ is the $j$-th parameter.  Within our moment approximation framework, the $s_{ij}$'s may either be estimated by perturbing the model's parameters and computing a finite difference approximation, or obtained automatically by employing the CVODES solver of \citet{Serban2003} when solving the system of ODEs (Equations \eqref{eq6} and \eqref{cmgf}).  
\afterpage{
\begin{figure*}[!ht]
\includegraphics[width = \linewidth,trim=0.05cm 0.00cm 0.05cm 0.05cm, clip=true]{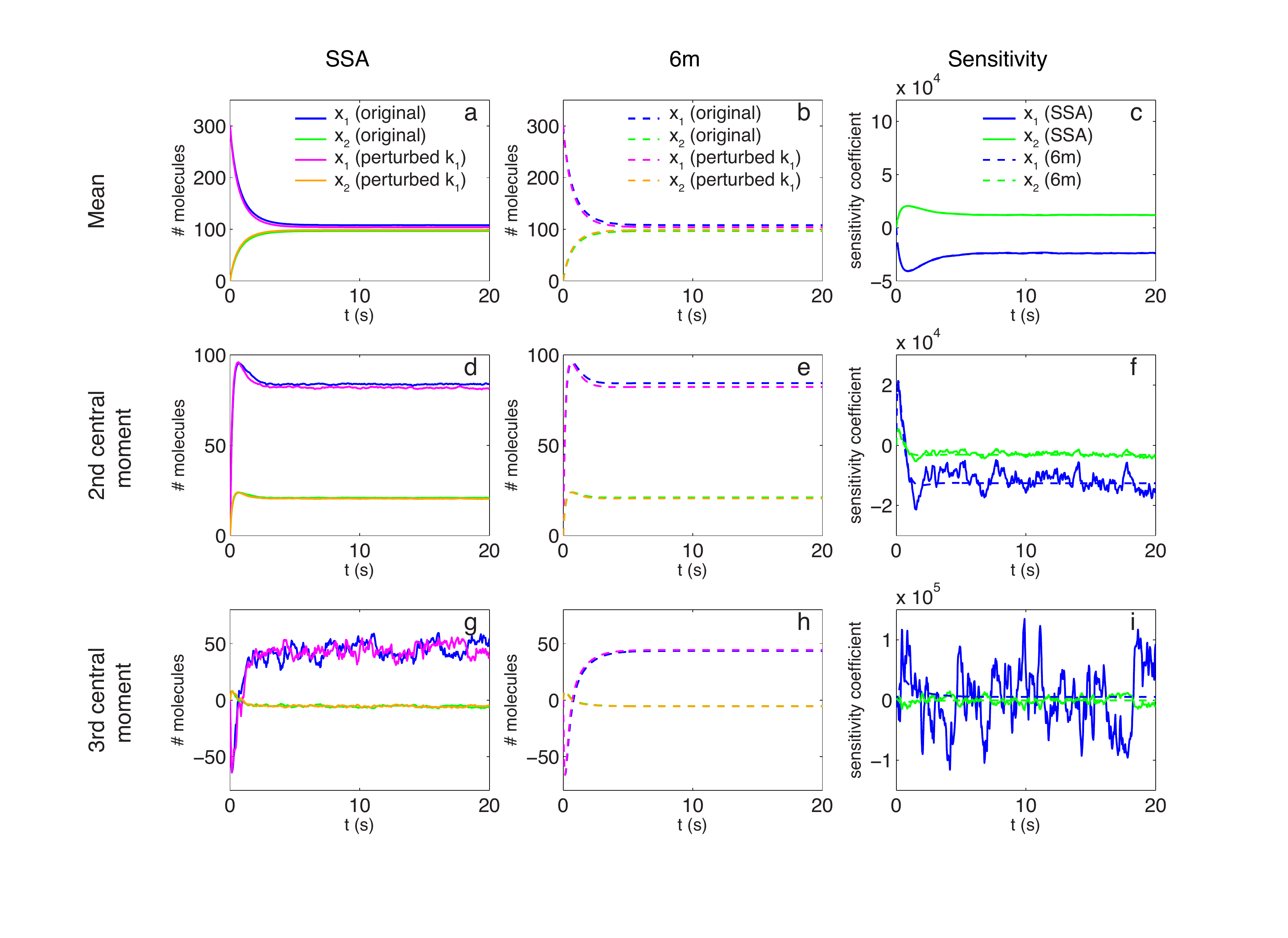}\caption{Assessing the sensitivity to parameter $k_1$ of the dimerisation system of Section III A.  Initial conditions and parameters are the same as previously, except that we additionally consider a perturbed value of $k_1 = 1.1 \times 1.66\cdot10^{-3}$   a) Average of 100,000 SSA simulations for the original and perturbed $k_1$ values.  b) As (a), except the mean is estimated using the proposed moment expansion approximation with 6 moments.  c) The sensitivity coefficients ${\partial \mu_1(t, [k_1, k_2])}/{\partial k_1}$ and ${\partial \mu_2(t, [k_1, k_2])}/{\partial k_2}$ (where $\mu_1$ and $\mu_2$ represent the means of $x_1$ and $x_2$ respectively) estimated from the SSA and 6m output via finite different approximations.  d), e) and f) As a), b) and c), but for the second central moment.   g), h) and i) As a), b) and c), but for the third central moment.   }\label{sens-fig01}
\end{figure*}
}
In Figure \ref{sens-fig01}, we reconsider the dimerisation model of Section III A.  We focus upon the sensitivity of the mean and  2nd and 3rd central moments of the two molecular species to the parameter $k_1$ (similar results are obtained for the sensitivity to $k_2$).  Figures \ref{sens-fig01}a, d and g show how the moments estimated from 100,000 SSA runs vary when we increase the original value of $k_1$ by 10 percent.  Figures \ref{sens-fig01}b, e and h show the same for the moments estimated using our proposed approach with 6 central moments (6m).  Figures \ref{sens-fig01}c, f and i show sensitivity coefficients estimated from both the SSA and 6m outputs using a finite difference approximation (in the 6m case, the sensitivity coefficients may instead be obtained automatically using the CVODES solver, which yields identical results).  There is generally good agreement between the coefficients estimated using the two different approaches.  However, as we consider higher moments, our ability to assess sensitivity using the SSA output rapidly diminishes, since the variability caused by the change in the parameter value is overwhelmed by the variability in the estimator due to finite sample size.  This may be rectified by increasing the number of SSA simulations, but at considerable computational cost.  In contrast, the sensitivity coefficients associated with higher moments may still be straightforwardly calculated using the moment expansion approach (although, of course, care must be taken to ensure that appropriately many moments have been taken into account by the approximation --- see Section III E).

\subsection{Simple Heuristics for Moment Expansions}
\begin{figure}
\includegraphics[width=0.48\textwidth]{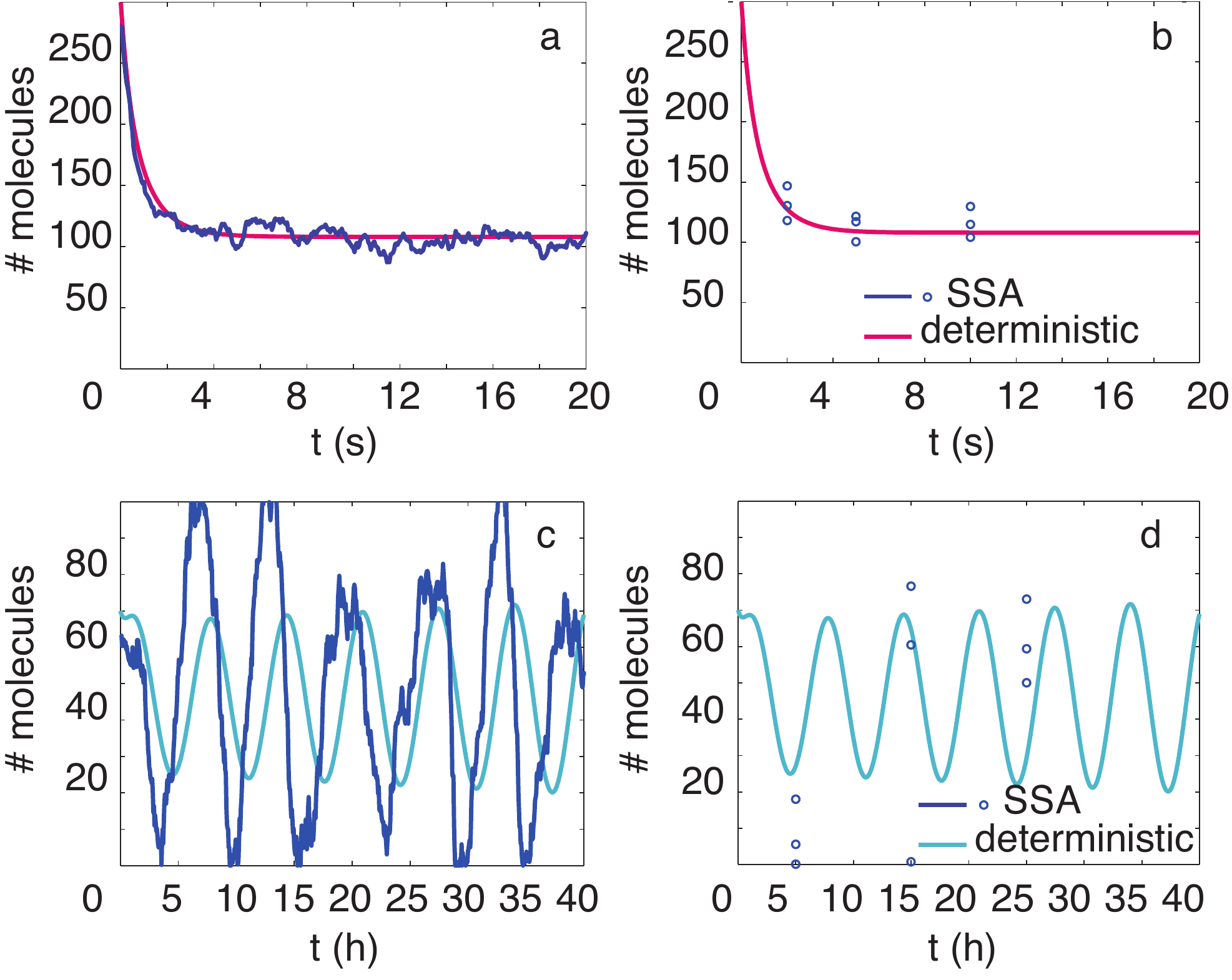}
\caption{\label{}Study of the deviation from the mean $(x-\mu)$ (a) Deterministic mean and single SSA trajectory for dimerisation system. (b) Deterministic mean and 9 points taken from different SSA trajectories for dimerisation system. (c) Deterministic mean and single SSA trajectory for p53 system. (d) Deterministic mean and 9 points taken from different SSA trajectories for p53 system. }
\label{p53figure5}
\end{figure}

Our results for the p53 system clearly demonstrate that failure to take a sufficient number of moments into account can lead to incorrect conclusions and biased parameter estimates.  Ideally we would like to know from the outset whether a deterministic approach or a two moment approximation is sufficient to capture the general statistical behaviour of the stochastic system.  But without recourse to a large number of SSA runs it is impossible to predict the statistical properties of the solutions to non-linear stochastic systems.  And in such cases it is generally not feasible to perform large numbers of SSA simulations and we need a different approach. We should look at the assumption made at the beginning of our derivation, where we assumed that we can approximate the propensity functions with a Taylor expansion. For a single variable a Taylor expansion of a function $f(x)$ about the point $c$ has the general form
\begin{eqnarray}
f(x)=f(c)+\frac{f'x}{\partial x}\left(x-c\right)+\frac{f''x}{\partial^2 x}\left(x-c\right)^2+\ldots
\end{eqnarray}
By taking into account only the first term, we assume that the function value in point $x$ is the same as for point $c$; by taking into account also the second term we assume that $f(x)$ can be approximated by a straight line between points $x$ and $c$, etc.. Truncating the expansion at a low order will only result in a good approximation in case $x$ is close to point $c$, where we have approximated the true function. In our case $c$ is the mean, $\mu$, implying that an approximation using a few moments will be accurate only in case all observations are close to the mean. In case it is possible to perform a single realisation of the SSA, we can assess this quality by comparing the mean calculated with the deterministic approach with the trajectory calculated with the SSA (Figure \ref{p53figure5}). 
\par
Figure \ref{p53figure5}a displays the deterministic mean of $x_1$ and one SSA simulation for the dimerisation system. In this case the trajectory is close to the mean over the complete time course. A single trajectory for p53, displayed in Figure \ref{p53figure5}c, compared to the deterministic result shows that in this case the distance of the trajectory from the mean is much larger. In case a single SSA realisation of the model is not possible, but experimental data are available, the distance form the mean can still be investigated in the same way. Figures \ref{p53figure5}b,d show the mean calculated using the deterministic approach together with `measured' values at three time points (obtained with the SSA), repeated three times, resulting in nine data points. Also when looking at the distance form these nine points from the deterministic mean, it is clear that for the dimerisation system $(x-\mu)$ is small, whereas for the p53 system it is {relatively} large, indicating that a larger number of moments is necessary to capture the full distribution.
\par
Such simple heuristics have the advantage of being computationally affordable. While inadequate at guaranteeing good performance of an expansion using any finite number of moments, we can use them to capture any gross inadequacy of a given approximation relatively reliably. Such small-scale analyses should precede or accompany moment expansions. More generally, we can consider this problem from the point of view of statistical model checking; see e.g. \cite{Gelman:2003}. But the question as to how many stochastic simulations need to be averaged over to get a good idea of the mean (or any higher moment) is challenging to answer for all but the most trivial systems \cite{Toni:2008aa}.

\subsection{Computational Complexity}
{The computational complexity of the moment approximation method depends on the number of variables in the system, and the number of terms that need to be evaluated for each central moment. Because of the symmetry of central moments, e.g.  $\left<(x_1-\mu_1)(x_2-\mu_2)\right>=\left<(x_2-\mu_2)(x_1-\mu_1)\right>$, there are many terms we do not need to include in the ODE representation of the system. The total number of central moment terms that could be nonzero when approximating a system with $d$ variables and up to $N$ moments is given by
\begin{eqnarray}
N_{cm}=\frac{(N+d)!}{N!d!}-d-1
\end{eqnarray}
We subtract $d$ terms because the first order central moments are always zero, and one term corresponding to the zeroth order central moment. For the deterministic case, the total number of ODE equations necessary to describe the system is equal to the number of variables, each term describes the mean of one of the variables. For the LNA the total number of equations is equal to the number of equations needed for the two moment approximation. We displayed the number of central moment terms to be evaluated for systems up to 9 variables and up to 6 central moments in Table \ref{table2}. When two central moments are included for 2 variables, we need to evaluate two ODE equations corresponding to the mean, and three equations for the second central moments, i.e. the variance of both variables, and their covariance. From Table \ref{table2} we can see that the number of equations that need to be included rises quickly with the number of variables in the system. 
\par
However, in order to obtain the same information on higher order moments by simulation with the SSA, a large number of simulation runs would have to be performed; in our experience even for relatively simple systems this is of the order of $>10^4-10^6$. We have already seen from the examples described above that the number of simulations necessary to obtain a smooth estimate of the higher order central moments increases with each order. For example, the third central moment in the dimerisation system calculated from 100,000 simulations still displays fluctuations that can only be expected to smooth out when more simulations were performed. }
\begin{table}
 \caption{\label{table2}Number of potentially nonzero central moment terms to include in moment approximation for different numbers of variables (columns) and number of central moments (rows).}
 
\vspace{5mm}
\setlength{\tabcolsep}{4.5pt}
\begin{tabular}{ l | c | c | c | c | c | c | c | c  c }
\hline \hline
\multicolumn{9}{c}{N variables} \\
\hline
 $ $& 2 & 3 & 4 & 5 & 6 & 7 & 8 & 9\\
 \hline 
$2m$ & 3    &  6 & 10& 15 & 21&28 & 36& 45\\
 $3m$ & 7   &  16 & 30& 50 &77 & 112& 156& 210\\
 $4m$ &  12 &  31& 65 &  120&203&322& 486& 705\\
 $5m$ &  18 &  52&  121& 246 &455 &784& 1278& 1992\\
 $6m$ &  25 &  80 &  205 & 456 & 917 &  1708& 2994& 4995\\
 \hline \hline
\end{tabular}
\end{table}

\section{Conclusion}
In this paper we have described a general moment expansion method for approximating the time evolution of stochastic kinetic systems. We have shown that taking into account more moments improves the estimate of the mean and the higher order moments. In case the deterministic approach delivers an accurate estimation of the mean, expanding the higher order moments still gives additional information on the variance, skewness and kurtosis of the distribution of the variables. Even for very simple systems, e.g. the dimerisation system, we find that higher moments obtained from averaging over 100,000 SSA runs are still fluctuating noticeably. 
\par
Instead of performing large numbers of SSA simulations, the time evolution of higher order moments may be obtained much more efficiently from moment approximation methods. Unfortunately it is not possible to predict {\em a priori} how many moments are required to fully capture the output distributions of stochastic processes. And while our method can be used to evaluate arbitrary higher moments, the number of equations that need to be solved increases rapidly with the number of molecular species and the number of moments required. This comes with increased computational cost for calculating the expressions as well as for integrating the equations. Therefore it is desirable to identify whether more moments are needed to model the system, and we have provided a simple heuristic to achieve this in many cases. 
While this can indicate if more than two moments are needed, it will not clearly identify when sufficiently many central moments have been included in the expansion. This is a point for further future investigation. In cases where the distribution is known, but different from normal, knowledge about the distribution can be used to close the set of equations correspondingly. The estimated higher order moments may also by themselves indicate which distribution is a probable candidate for the output of a given dynamical stochastic system; we can, for example, compare obtained estimates of such moments against known values for the higher order moments for particular distributions: if e.g. the estimated skewness and kurtosis are  close to zero and three, respectively, a normal distribution might be a good choice and an approximation similar to the simple linear noise approximation may be appropriate.
\par
When the model under investigation is non-linear, lower order moments typically depend on higher order moments, and the system of equations is not closed. Closure can be achieved in several different ways. Here we have closed the system by truncating the Taylor series expansion at a selected order. Several other approaches have been previously evaluated. Chevalier et al \cite{Chevalier2011}, for example, approximated higher order moments using experimental data. Azunre et al\cite{Azunre2011} showed that for very small molecule numbers, using only two moments can lead to unstable results. Singh et al. \cite{Singh2011} developed a derivative-matching approach in the context of polynomial rate equations, which proved to work very well in particular for small molecule numbers. {For the examples described in this paper, the proposed method shows differences in behaviour for very low molecule numbers. More specifically, for the dimerisation system, the mean trajectories become unstable below approximately 7 molecules in the system; for the Michaelis-Menten example, the system does not become unstable, but becomes less accurate for very small molecule numbers. The p53 system trajectories for the system size as displayed in Fig. \ref{p53figure1} start running out of phase around $t=35$, and when the molecule number decreases, the point where the trajectories run out of phase moves forward, followed by instability in the trajectories. The LNA solution for the p53 system does not become unstable for small molecule numbers but keeps showing the expanding oscillatory behaviour.}
For other systems that show periodic behaviour it might be more beneficial to approximate the system in the frequency domain. We would recommend use of this method --- certainly if no manual closure is attempted, but probably even then --- for moderate and large numbers of molecules; a conservative estimate might be to have at least 10-20 molecules. However, for molecular abundances of less than 10 molecules we find that the numerical burden of the SSA compared to the MFK approaches is no longer prohibitive, and, again conservatively, we would consider the use of the SSA.
\par
The present work also provides a natural framework for an inferential framework: our results above suggest that inclusion of higher order moments will lead to increased accuracy of the parameter estimates. Milner et al. \cite{Milner2011b} derived a likelihood function that included the mean and variance obtained from a moment closure method, assuming a multivariate Gaussian distribution. Kuegler \cite{Kuegler2012} used both the mean and the variance obtained from a moment closure method for parameter estimation by minimizing an objective function that included both the difference between the observed and estimated mean and the difference between the observed and estimated variance. This showed that more accurate parameter estimates could be obtained when the observed variance is included for parameter estimation. Other analyses have employed approximate Bayesian computation schemes \cite{Toni:2012bo}, {and used moment-based inference employing the mean and variance to infer parameters of a bimodal system\cite{Zechner2012}}. Through recent technological advances in the field of single cell observations \cite{Klug2011,Lin2011,Ozaki:2010p26116}, it becomes possible to probe directly the properties of the output distributions of stochastic dynamical systems over time. The additional information about the higher order central moments that can be derived from these datasets can be exploited when higher  order empirical moments are also used for parameter inference.  However, while likelihoods are trivially constructed when the mean behaviour and variance estimates are available {(via Gaussian assumptions)}, conditioning on higher-order moments typically requires some further assumptions; most easily these moments are included in maximum-entropy estimators of the probability distribution. The present approach yields these moments, however, reliably and affordably. Because of their affordability these moments also open up new ways for assessing the sensitivity of stochastic dynamical systems (as outlined in section IIID), including cases where the linear noise approximation tends to break down \cite{Komorowski2010}. This includes general feedback systems where the notion of sensitivity may be particularly useful but calculation for stochastic systems is fraught with problems and numerically expensive. In conclusion, the general moment expansion method described in this paper provides a flexible and valuable new tool for investigating many stochastic kinetic systems.



%
%

%



\appendix

\section{Model equations}
{In this section we display the model equations used for the dimerisation and Michaelis-Menten system. The complexity of the equations grows with the number of moments included, we display here the equations used for the mean, variance and co-variance when truncating after second order.}
\subsection{Dimerisation}
For the dimerisation system, the equations used for the mean, variance and covariance are given by
\begin{align}
&\mu_{x_1}=2k_2x_2-2k_1\sigma^2_{x_1^2}-2k_1x_1(x_1-1)\nonumber\\
&\mu_{x_2}=k_1\sigma^2_{x_1^2}-k_2x_2+k_1x_1(x_1-1)\nonumber\\
&\sigma^2_{x_2^2}=k_1y_{1}^2-k_1x_1+k_2x_2+c1\sigma^2_{x_1^2}-2k_2\sigma^2_{x_2^2}-2k_1\sigma^2_{x_1,x_2}\nonumber\\
& \  \  \  \  \  \  \  \ +4k_1x_1\sigma^2_{x_1,x_2}\nonumber\\
&\sigma^2_{x_1,x_2}=2k_1\sigma^2_{x_1,x_2}+2k_2\sigma^2_{x_2^2}-k_1\sigma^2_{x_1^2}-k_2\sigma^2_{x_1,x_2}-2k_1x_1^2\nonumber\\
& \  \  \  \  \  \  \  \ -2k_1x_1+2k_2x_2+2k_1\sigma^2_{x_1^2}-4k_1x_1\sigma^2_{x_1,x_2}+2k_1x_1\sigma^2_{x_1^2}\nonumber\\
&\sigma^2_{x_1^2}=4k_1\sigma^2_{x_1^2}+4k_2\sigma^2_{x_1,x_2}+4k_1x_1^2-4k_1x_1+4k_2x_2+\nonumber\\
& \  \  \  \  \  \  \  \ 4k_1\sigma^2_{x_1^2}-8k_1x_1\sigma^2_{x_1^2}\nonumber
\end{align}

\subsection{Michaelis-Menten System}
For the Michaelis-Menten system, the equations used for the mean, variance and covariance are given by
\begin{align}
&\mu_{x_1}=-k_2(x_1+x_2-301)-k_1\sigma^2_{x_1,x_2}-\nonumber\\
& \  \  \  \  \  \  \  \ k_1 \sigma^2_{x_1^2}-k_1x_1(x_1+x_2-181)\nonumber\\
&\mu_{x_2}=-c_3(x_1+x_2-301)\nonumber\\
&\sigma^2_{x_2^2}=-2c_3(\sigma^2_{x_2^2}+\sigma^2_{x_1,x_2})-c_3(x_1+x_2-301)\nonumber\\
&\sigma^2_{x_1,x_2}=181k_1\sigma^2_{x_1,x_2}-k_2\sigma^2_{x_2^2}-k_2\sigma^2_{x_1,x_2}-c_3\sigma^2_{x_1,x_2}\nonumber\\
& \  \  \  \  \  \  \  \ -c_3\sigma^2_{x_1^2}-k_1x_1\sigma^2_{x_2^2}-2k_1x_1\sigma^2_{x_1,x_2}-k_1x_2\sigma^2_{x_1,x_2}\nonumber\\
&\sigma^2_{x_1^2}=-(181k_1x_1-301k_2+k_2x_1+k_2x_2\nonumber\\
& \  \  \  \  \  \  \  \ -k_1\sigma^2_{x_1,x_2}-k_1\sigma^2_{x_1^2}-k_1x_1^2-362k_1\sigma^2_{x_1^2}+\nonumber\\
& \  \  \  \  \  \   \  \ 2k_2\sigma^2_{x_1,x_2}+2k_2\sigma^2_{x_1^2}-k_1x_1x_2+2k_1x_1\sigma^2_{x_1,x_2}\nonumber\\
& \  \  \  \  \  \  \  \ +4k_1x_1\sigma^2_{x_1^2}+2k_1x_2\sigma^2_{x_1^2})\nonumber
\end{align}


\begin{thebibliography}{46}%
\makeatletter
\providecommand \@ifxundefined [1]{%
 \@ifx{#1\undefined}
}%
\providecommand \@ifnum [1]{%
 \ifnum #1\expandafter \@firstoftwo
 \else \expandafter \@secondoftwo
 \fi
}%
\providecommand \@ifx [1]{%
 \ifx #1\expandafter \@firstoftwo
 \else \expandafter \@secondoftwo
 \fi
}%
\providecommand \natexlab [1]{#1}%
\providecommand \enquote  [1]{``#1''}%
\providecommand \bibnamefont  [1]{#1}%
\providecommand \bibfnamefont [1]{#1}%
\providecommand \citenamefont [1]{#1}%
\providecommand \href@noop [0]{\@secondoftwo}%
\providecommand \href [0]{\begingroup \@sanitize@url \@href}%
\providecommand \@href[1]{\@@startlink{#1}\@@href}%
\providecommand \@@href[1]{\endgroup#1\@@endlink}%
\providecommand \@sanitize@url [0]{\catcode `\\12\catcode `\$12\catcode
  `\&12\catcode `\#12\catcode `\^12\catcode `\_12\catcode `\%12\relax}%
\providecommand \@@startlink[1]{}%
\providecommand \@@endlink[0]{}%
\providecommand \url  [0]{\begingroup\@sanitize@url \@url }%
\providecommand \@url [1]{\endgroup\@href {#1}{\urlprefix }}%
\providecommand \urlprefix  [0]{URL }%
\providecommand \Eprint [0]{\href }%
\providecommand \doibase [0]{http://dx.doi.org/}%
\providecommand \selectlanguage [0]{\@gobble}%
\providecommand \bibinfo  [0]{\@secondoftwo}%
\providecommand \bibfield  [0]{\@secondoftwo}%
\providecommand \translation [1]{[#1]}%
\providecommand \BibitemOpen [0]{}%
\providecommand \bibitemStop [0]{}%
\providecommand \bibitemNoStop [0]{.\EOS\space}%
\providecommand \EOS [0]{\spacefactor3000\relax}%
\providecommand \BibitemShut  [1]{\csname bibitem#1\endcsname}%
\let\auto@bib@innerbib\@empty
\bibitem [{\citenamefont {Gardiner}(2009)}]{Gardiner2009}%
  \BibitemOpen
  \bibfield  {author} {\bibinfo {author} {\bibfnamefont {C.}~\bibnamefont
  {Gardiner}},\ }\href@noop {} {\emph {\bibinfo {title} {{Stochastic
  methods}}}},\ \bibinfo {edition} {4th}\ ed.,\ Springer Series in Synergetics\
  (\bibinfo  {publisher} {Springer-Verlag},\ \bibinfo {address} {Berlin},\
  \bibinfo {year} {2009})\BibitemShut {NoStop}%
\bibitem [{\citenamefont {van Kampen}(1992)}]{Kampen:1992aa}%
  \BibitemOpen
  \bibfield  {author} {\bibinfo {author} {\bibfnamefont {N.}~\bibnamefont {van
  Kampen}},\ }\href@noop {} {\emph {\bibinfo {title} {Stochastic Processes in
  Physics and Chemistry}}}\ (\bibinfo  {publisher} {North-Holland},\ \bibinfo
  {year} {1992})\BibitemShut {NoStop}%
\bibitem [{\citenamefont {{Gillespie}}(1976)}]{Gillespie1976}%
  \BibitemOpen
  \bibfield  {author} {\bibinfo {author} {\bibfnamefont {D.~T.}\ \bibnamefont
  {{Gillespie}}},\ }\bibfield  {title} {\enquote {\bibinfo {title} {General
  method for numerically simulating stochastic time evolution of coupled
  chemical-reactions},}\ }\href@noop {} {\bibfield  {journal} {\bibinfo
  {journal} {Journal of Computational Physics}\ }\textbf {\bibinfo {volume}
  {22}},\ \bibinfo {pages} {403--434} (\bibinfo {year} {1976})}\BibitemShut
  {NoStop}%
\bibitem [{\citenamefont {{Gillespie}}(2009)}]{Gillespie2009}%
  \BibitemOpen
  \bibfield  {author} {\bibinfo {author} {\bibfnamefont {C.~S.}\ \bibnamefont
  {{Gillespie}}},\ }\bibfield  {title} {\enquote {\bibinfo {title}
  {Moment-closure approximations for mass-action models},}\ }\href@noop {}
  {\bibfield  {journal} {\bibinfo  {journal} {The Institution of Engineering
  and Technology}\ }\textbf {\bibinfo {volume} {3}},\ \bibinfo {pages} {52--58}
  (\bibinfo {year} {2009})}\BibitemShut {NoStop}%
\bibitem [{\citenamefont {{G\`omez-Uribe}}\ and\ \citenamefont
  {{Verghese}}(2007)}]{Gomez2007}%
  \BibitemOpen
  \bibfield  {author} {\bibinfo {author} {\bibfnamefont {C.~A.}\ \bibnamefont
  {{G\`omez-Uribe}}}\ and\ \bibinfo {author} {\bibfnamefont {G.~C.}\
  \bibnamefont {{Verghese}}},\ }\bibfield  {title} {\enquote {\bibinfo {title}
  {Mass fluctuation kinetics: capturing stochastic effects in systems of
  chemical reactions through coupled mean-variance computations},}\ }\href@noop
  {} {\bibfield  {journal} {\bibinfo  {journal} {Journal of Chemical Physics}\
  }\textbf {\bibinfo {volume} {126}},\ \bibinfo {pages} {024109--024112}
  (\bibinfo {year} {2007})}\BibitemShut {NoStop}%
\bibitem [{\citenamefont {{Lee}}, \citenamefont {{Kim}},\ and\ \citenamefont
  {{Kim}}(2009)}]{Lee2009}%
  \BibitemOpen
  \bibfield  {author} {\bibinfo {author} {\bibfnamefont {C.}~\bibnamefont
  {{Lee}}}, \bibinfo {author} {\bibfnamefont {K.-H.}\ \bibnamefont {{Kim}}}, \
  and\ \bibinfo {author} {\bibfnamefont {P.}~\bibnamefont {{Kim}}},\ }\bibfield
   {title} {\enquote {\bibinfo {title} {A moment closure method for stochastic
  reaction networks},}\ }\href@noop {} {\bibfield  {journal} {\bibinfo
  {journal} {The Journal of Chemical Physics}\ }\textbf {\bibinfo {volume}
  {130}},\ \bibinfo {pages} {134107} (\bibinfo {year} {2009})}\BibitemShut
  {NoStop}%
\bibitem [{\citenamefont {{Singh}}\ and\ \citenamefont
  {{Hespanha}}(2010)}]{Singh2010}%
  \BibitemOpen
  \bibfield  {author} {\bibinfo {author} {\bibfnamefont {A.}~\bibnamefont
  {{Singh}}}\ and\ \bibinfo {author} {\bibfnamefont {J.}~\bibnamefont
  {{Hespanha}}},\ }\bibfield  {title} {\enquote {\bibinfo {title} {Stochastic
  hybrid systems for studying biochemical processes},}\ }\href@noop {}
  {\bibfield  {journal} {\bibinfo  {journal} {Phil. Trans. R. Soc. A}\ }\textbf
  {\bibinfo {volume} {368}},\ \bibinfo {pages} {4995--5011} (\bibinfo {year}
  {2010})}\BibitemShut {NoStop}%
\bibitem [{\citenamefont {{Singh}}\ and\ \citenamefont
  {{Hespanha}}(2006)}]{Singh2006}%
  \BibitemOpen
  \bibfield  {author} {\bibinfo {author} {\bibfnamefont {A.}~\bibnamefont
  {{Singh}}}\ and\ \bibinfo {author} {\bibfnamefont {J.}~\bibnamefont
  {{Hespanha}}},\ }\bibfield  {title} {\enquote {\bibinfo {title} {Lognormal
  moment closures for biochemical reactions},}\ }in\ \href@noop {} {\emph
  {\bibinfo {booktitle} {Decision and Control 2006 45th IEEE Conference}}}\
  (\bibinfo {year} {2006})\ p.\ \bibinfo {pages} {2063Ð2068}\BibitemShut
  {NoStop}%
\bibitem [{\citenamefont {{Ruess}}\ \emph {et~al.}(2011)\citenamefont
  {{Ruess}}, \citenamefont {{Milias-Argeitis}}, \citenamefont {{Summers}},\
  and\ \citenamefont {{Lygeros}}}]{Ruess2011}%
  \BibitemOpen
  \bibfield  {author} {\bibinfo {author} {\bibfnamefont {J.}~\bibnamefont
  {{Ruess}}}, \bibinfo {author} {\bibfnamefont {A.}~\bibnamefont
  {{Milias-Argeitis}}}, \bibinfo {author} {\bibfnamefont {S.}~\bibnamefont
  {{Summers}}}, \ and\ \bibinfo {author} {\bibfnamefont {J.}~\bibnamefont
  {{Lygeros}}},\ }\bibfield  {title} {\enquote {\bibinfo {title} {Moment
  estimation for chemically reacting systems by extended kalman filtering},}\
  }\href@noop {} {\bibfield  {journal} {\bibinfo  {journal} {The Journal of
  Chemical Physics}\ }\textbf {\bibinfo {volume} {135}},\ \bibinfo {pages}
  {165102} (\bibinfo {year} {2011})}\BibitemShut {NoStop}%
\bibitem [{\citenamefont {{Goutsias}}(2007)}]{Goutsias2007}%
  \BibitemOpen
  \bibfield  {author} {\bibinfo {author} {\bibfnamefont {J.}~\bibnamefont
  {{Goutsias}}},\ }\bibfield  {title} {\enquote {\bibinfo {title} {Classical
  versus stochastic kinetics modeling of biochemical reaction systems},}\
  }\href@noop {} {\bibfield  {journal} {\bibinfo  {journal} {Biophysical
  Journal}\ }\textbf {\bibinfo {volume} {92}},\ \bibinfo {pages} {2350--2365}
  (\bibinfo {year} {2007})}\BibitemShut {NoStop}%
\bibitem [{\citenamefont {{Barzel}}\ and\ \citenamefont
  {{Biham}}(2012)}]{Barzel2012}%
  \BibitemOpen
  \bibfield  {author} {\bibinfo {author} {\bibfnamefont {B.}~\bibnamefont
  {{Barzel}}}\ and\ \bibinfo {author} {\bibfnamefont {O.}~\bibnamefont
  {{Biham}}},\ }\bibfield  {title} {\enquote {\bibinfo {title} {Stochastic
  analysis of complex reaction networks using binomial moment equations},}\
  }\href@noop {} {\bibfield  {journal} {\bibinfo  {journal} {Physical Review
  E}\ }\textbf {\bibinfo {volume} {86}},\ \bibinfo {pages} {031126} (\bibinfo
  {year} {2012})}\BibitemShut {NoStop}%
\bibitem [{\citenamefont {{Komorowski}}\ \emph {et~al.}(2009)\citenamefont
  {{Komorowski}}, \citenamefont {{Finkenstaedt}}, \citenamefont {{Harper}},\
  and\ \citenamefont {{Rand}}}]{Komorowski2009}%
  \BibitemOpen
  \bibfield  {author} {\bibinfo {author} {\bibfnamefont {M.}~\bibnamefont
  {{Komorowski}}}, \bibinfo {author} {\bibfnamefont {B.}~\bibnamefont
  {{Finkenstaedt}}}, \bibinfo {author} {\bibfnamefont {C.~V.}\ \bibnamefont
  {{Harper}}}, \ and\ \bibinfo {author} {\bibfnamefont {D.~A.}\ \bibnamefont
  {{Rand}}},\ }\bibfield  {title} {\enquote {\bibinfo {title} {Bayesian
  inference of biochemical kinetic parameters using the linear noise
  approximation},}\ }\href@noop {} {\bibfield  {journal} {\bibinfo  {journal}
  {BMC bioinformatics}\ }\textbf {\bibinfo {volume} {10}} (\bibinfo {year}
  {2009})}\BibitemShut {NoStop}%
\bibitem [{\citenamefont {{Komorowski}}\ \emph {et~al.}(2011)\citenamefont
  {{Komorowski}}, \citenamefont {{Costa}}, \citenamefont {{Rand}},\ and\
  \citenamefont {{Stumpf}}}]{Komorowski2010}%
  \BibitemOpen
  \bibfield  {author} {\bibinfo {author} {\bibfnamefont {M.}~\bibnamefont
  {{Komorowski}}}, \bibinfo {author} {\bibfnamefont {M.~J.}\ \bibnamefont
  {{Costa}}}, \bibinfo {author} {\bibfnamefont {D.~A.}\ \bibnamefont {{Rand}}},
  \ and\ \bibinfo {author} {\bibfnamefont {M.~P.~H.}\ \bibnamefont
  {{Stumpf}}},\ }\bibfield  {title} {\enquote {\bibinfo {title} {Sensitivity,
  robustness, and identifiability in stochastic chemical kinetics models},}\
  }\href@noop {} {\bibfield  {journal} {\bibinfo  {journal} {PNAS}\ }\textbf
  {\bibinfo {volume} {108}},\ \bibinfo {pages} {8645--50} (\bibinfo {year}
  {2011})}\BibitemShut {NoStop}%
\bibitem [{\citenamefont {{Pahle}}\ \emph {et~al.}(2012)\citenamefont
  {{Pahle}}, \citenamefont {{Challenger}}, \citenamefont {{Mendes}},\ and\
  \citenamefont {{McKane}}}]{Pahle2012}%
  \BibitemOpen
  \bibfield  {author} {\bibinfo {author} {\bibfnamefont {J.}~\bibnamefont
  {{Pahle}}}, \bibinfo {author} {\bibfnamefont {J.}~\bibnamefont
  {{Challenger}}}, \bibinfo {author} {\bibfnamefont {P.}~\bibnamefont
  {{Mendes}}}, \ and\ \bibinfo {author} {\bibfnamefont {A.}~\bibnamefont
  {{McKane}}},\ }\bibfield  {title} {\enquote {\bibinfo {title} {Biochemical
  fluctuations, optimisation and the linear noise approximation},}\ }\href@noop
  {} {\bibfield  {journal} {\bibinfo  {journal} {BMC Systems Biology}\ }\textbf
  {\bibinfo {volume} {86}} (\bibinfo {year} {2012})}\BibitemShut {NoStop}%
\bibitem [{\citenamefont {{Wallace}}\ \emph {et~al.}(2012)\citenamefont
  {{Wallace}}, \citenamefont {{Gillespie}}, \citenamefont {{Sanft}},\ and\
  \citenamefont {{Petzold}}}]{Wallace2012}%
  \BibitemOpen
  \bibfield  {author} {\bibinfo {author} {\bibfnamefont {E.}~\bibnamefont
  {{Wallace}}}, \bibinfo {author} {\bibfnamefont {D.}~\bibnamefont
  {{Gillespie}}}, \bibinfo {author} {\bibfnamefont {K.}~\bibnamefont
  {{Sanft}}}, \ and\ \bibinfo {author} {\bibfnamefont {L.}~\bibnamefont
  {{Petzold}}},\ }\bibfield  {title} {\enquote {\bibinfo {title} {Linear noise
  approximation is valid over limited times for any chemical system that is
  sufficiently large},}\ }\href@noop {} {\bibfield  {journal} {\bibinfo
  {journal} {IET Syst. Biol.}\ }\textbf {\bibinfo {volume} {6}},\ \bibinfo
  {pages} {102--115} (\bibinfo {year} {2012})}\BibitemShut {NoStop}%
\bibitem [{\citenamefont {{Singh}}\ and\ \citenamefont
  {{Hespanha}}(2011)}]{Singh2011}%
  \BibitemOpen
  \bibfield  {author} {\bibinfo {author} {\bibfnamefont {A.}~\bibnamefont
  {{Singh}}}\ and\ \bibinfo {author} {\bibfnamefont {J.~P.}\ \bibnamefont
  {{Hespanha}}},\ }\bibfield  {title} {\enquote {\bibinfo {title} {Approximate
  moment dynamics for chemically reacting systems},}\ }\href@noop {} {\bibfield
   {journal} {\bibinfo  {journal} {IEEE Transactions on Automatic Control}\
  }\textbf {\bibinfo {volume} {56}},\ \bibinfo {pages} {414--418} (\bibinfo
  {year} {2011})}\BibitemShut {NoStop}%
\bibitem [{\citenamefont {{Sotiropoulos}}\ and\ \citenamefont
  {{Kaznessis}}(2011)}]{Sotiropoulos2011}%
  \BibitemOpen
  \bibfield  {author} {\bibinfo {author} {\bibfnamefont {V.}~\bibnamefont
  {{Sotiropoulos}}}\ and\ \bibinfo {author} {\bibfnamefont {Y.~N.}\
  \bibnamefont {{Kaznessis}}},\ }\bibfield  {title} {\enquote {\bibinfo {title}
  {Analytical derivation of moment equations in stochastic chemical
  kinetics},}\ }\href@noop {} {\bibfield  {journal} {\bibinfo  {journal}
  {Chemical Engineering Science}\ }\textbf {\bibinfo {volume} {66}},\ \bibinfo
  {pages} {268--277} (\bibinfo {year} {2011})}\BibitemShut {NoStop}%
\bibitem [{\citenamefont {{Milner}}, \citenamefont {{Gilespie}},\ and\
  \citenamefont {{Wilkinson}}(2011)}]{Milner2011}%
  \BibitemOpen
  \bibfield  {author} {\bibinfo {author} {\bibfnamefont {P.}~\bibnamefont
  {{Milner}}}, \bibinfo {author} {\bibfnamefont {C.~C.}\ \bibnamefont
  {{Gilespie}}}, \ and\ \bibinfo {author} {\bibfnamefont {D.~J.}\ \bibnamefont
  {{Wilkinson}}},\ }\bibfield  {title} {\enquote {\bibinfo {title} {Moment
  closure approximations for stochastic kinetic models with rational rate
  laws},}\ }\href@noop {} {\bibfield  {journal} {\bibinfo  {journal}
  {Mathematical Biosciences}\ }\textbf {\bibinfo {volume} {231}},\ \bibinfo
  {pages} {99--104} (\bibinfo {year} {2011})}\BibitemShut {NoStop}%
\bibitem [{\citenamefont {{Ullah}}\ and\ \citenamefont
  {{Wolkenhauer}}(2009)}]{Ullah2009}%
  \BibitemOpen
  \bibfield  {author} {\bibinfo {author} {\bibfnamefont {M.}~\bibnamefont
  {{Ullah}}}\ and\ \bibinfo {author} {\bibfnamefont {O.}~\bibnamefont
  {{Wolkenhauer}}},\ }\bibfield  {title} {\enquote {\bibinfo {title}
  {Investigating the two-moment characterisation of subcellular biochemical
  networks},}\ }\href@noop {} {\bibfield  {journal} {\bibinfo  {journal}
  {Journal of Theoretical Biology}\ }\textbf {\bibinfo {volume} {260}},\
  \bibinfo {pages} {340--352} (\bibinfo {year} {2009})}\BibitemShut {NoStop}%
\bibitem [{\citenamefont {{Grima}}(2012)}]{Grima2012}%
  \BibitemOpen
  \bibfield  {author} {\bibinfo {author} {\bibfnamefont {R.}~\bibnamefont
  {{Grima}}},\ }\bibfield  {title} {\enquote {\bibinfo {title} {General method
  for numerically simulating stochastic time evolution of coupled
  chemical-reactions},}\ }\href@noop {} {\bibfield  {journal} {\bibinfo
  {journal} {Journal of Computational Physics}\ }\textbf {\bibinfo {volume}
  {22}},\ \bibinfo {pages} {403--434} (\bibinfo {year} {2012})}\BibitemShut
  {NoStop}%
\bibitem [{\citenamefont {Batchelor}, \citenamefont {Loewer},\ and\
  \citenamefont {Lahav}(2009)}]{Batchelor:2009hk}%
  \BibitemOpen
  \bibfield  {author} {\bibinfo {author} {\bibfnamefont {E.}~\bibnamefont
  {Batchelor}}, \bibinfo {author} {\bibfnamefont {A.}~\bibnamefont {Loewer}}, \
  and\ \bibinfo {author} {\bibfnamefont {G.}~\bibnamefont {Lahav}},\ }\bibfield
   {title} {\enquote {\bibinfo {title} {{The ups and downs of p53:
  understanding protein dynamics in single cells}},}\ }\href@noop {} {\bibfield
   {journal} {\bibinfo  {journal} {Nature Reviews Cancer}\ }\textbf {\bibinfo
  {volume} {9}},\ \bibinfo {pages} {371--377} (\bibinfo {year}
  {2009})}\BibitemShut {NoStop}%
\bibitem [{\citenamefont {{Azunre}}(2007)}]{Azunre2007}%
  \BibitemOpen
  \bibfield  {author} {\bibinfo {author} {\bibfnamefont {P.}~\bibnamefont
  {{Azunre}}},\ }\emph {\bibinfo {title} {Mass Fluctuation Kinetics: Analysis
  and Computation of Equilibria and Local Dynamics}},\ \href@noop {} {Master's
  thesis},\ \bibinfo  {school} {Massachusetts Institute of Technology}
  (\bibinfo {year} {2007})\BibitemShut {NoStop}%
\bibitem [{\citenamefont {{Ito}}\ and\ \citenamefont
  {{Uchida}}(2010)}]{Ito2010}%
  \BibitemOpen
  \bibfield  {author} {\bibinfo {author} {\bibfnamefont {Y.}~\bibnamefont
  {{Ito}}}\ and\ \bibinfo {author} {\bibfnamefont {K.}~\bibnamefont
  {{Uchida}}},\ }\bibfield  {title} {\enquote {\bibinfo {title} {Formulas for
  intrinsic noise evaluation in oscillatory genetic networks},}\ }\href@noop {}
  {\bibfield  {journal} {\bibinfo  {journal} {Journal of theoretical biology}\
  }\textbf {\bibinfo {volume} {267}},\ \bibinfo {pages} {223--234} (\bibinfo
  {year} {2010})}\BibitemShut {NoStop}%
\bibitem [{\citenamefont {{Komorowski}}, \citenamefont {{Miekisz}},\ and\
  \citenamefont {{Stumpf}}(2013)}]{Komorowski2012}%
  \BibitemOpen
  \bibfield  {author} {\bibinfo {author} {\bibfnamefont {M.}~\bibnamefont
  {{Komorowski}}}, \bibinfo {author} {\bibfnamefont {J.}~\bibnamefont
  {{Miekisz}}}, \ and\ \bibinfo {author} {\bibfnamefont {M.~P.}\ \bibnamefont
  {{Stumpf}}},\ }\bibfield  {title} {\enquote {\bibinfo {title} {Decomposing
  noise in biochemical signalling systems highlights the role of protein
  degradation},}\ }\href@noop {} {\bibfield  {journal} {\bibinfo  {journal}
  {Biophysical Journal}\ }\textbf {\bibinfo {volume} {in press}} (\bibinfo
  {year} {2013})}\BibitemShut {NoStop}%
\bibitem [{\citenamefont {{Wilkinson}}(2012)}]{Wilkinson2012}%
  \BibitemOpen
  \bibfield  {author} {\bibinfo {author} {\bibfnamefont {D.~J.}\ \bibnamefont
  {{Wilkinson}}},\ }\href@noop {} {\emph {\bibinfo {title} {Stochastic
  Modelling for Systems Biology}}}\ (\bibinfo  {publisher} {Chapman and Hall,
  CRC press},\ \bibinfo {year} {2012})\BibitemShut {NoStop}%
\bibitem [{\citenamefont {{Geva-Zatorsky}}\ \emph {et~al.}(2006)\citenamefont
  {{Geva-Zatorsky}}, \citenamefont {{Rosenfeld}}, \citenamefont {{Itzkovitz}},
  \citenamefont {{Milo}}, \citenamefont {{Sigal}}, \citenamefont {{Dekel}},
  \citenamefont {{Yarnitzky}}, \citenamefont {{Liron}}, \citenamefont
  {{Polak}}, \citenamefont {{Lahav}},\ and\ \citenamefont {{Alon}}}]{Geva2006}%
  \BibitemOpen
  \bibfield  {author} {\bibinfo {author} {\bibfnamefont {N.}~\bibnamefont
  {{Geva-Zatorsky}}}, \bibinfo {author} {\bibfnamefont {N.}~\bibnamefont
  {{Rosenfeld}}}, \bibinfo {author} {\bibfnamefont {S.}~\bibnamefont
  {{Itzkovitz}}}, \bibinfo {author} {\bibfnamefont {R.}~\bibnamefont {{Milo}}},
  \bibinfo {author} {\bibfnamefont {A.}~\bibnamefont {{Sigal}}}, \bibinfo
  {author} {\bibfnamefont {E.}~\bibnamefont {{Dekel}}}, \bibinfo {author}
  {\bibfnamefont {T.}~\bibnamefont {{Yarnitzky}}}, \bibinfo {author}
  {\bibfnamefont {Y.}~\bibnamefont {{Liron}}}, \bibinfo {author} {\bibfnamefont
  {P.}~\bibnamefont {{Polak}}}, \bibinfo {author} {\bibfnamefont
  {G.}~\bibnamefont {{Lahav}}}, \ and\ \bibinfo {author} {\bibfnamefont
  {U.}~\bibnamefont {{Alon}}},\ }\bibfield  {title} {\enquote {\bibinfo {title}
  {Oscillations and variability in the p53 system},}\ }\href@noop {} {\bibfield
   {journal} {\bibinfo  {journal} {Molecular Systems Biology}\ }\textbf
  {\bibinfo {volume} {2}},\ \bibinfo {pages} {2006.0033} (\bibinfo {year}
  {2006})}\BibitemShut {NoStop}%
\bibitem [{\citenamefont {{Kirk}}, \citenamefont {{Toni}},\ and\ \citenamefont
  {{Stumpf}}(2008)}]{Kirk2008}%
  \BibitemOpen
  \bibfield  {author} {\bibinfo {author} {\bibfnamefont {P.~D.}\ \bibnamefont
  {{Kirk}}}, \bibinfo {author} {\bibfnamefont {T.}~\bibnamefont {{Toni}}}, \
  and\ \bibinfo {author} {\bibfnamefont {M.~P.}\ \bibnamefont {{Stumpf}}},\
  }\bibfield  {title} {\enquote {\bibinfo {title} {Parameter inference for
  biochemical systems that undergo a hopf bifurcation},}\ }\href@noop {}
  {\bibfield  {journal} {\bibinfo  {journal} {Biophysical Journal}\ }\textbf
  {\bibinfo {volume} {95}},\ \bibinfo {pages} {540Ð549} (\bibinfo {year}
  {2008})}\BibitemShut {NoStop}%
\bibitem [{\citenamefont {Saltelli}\ \emph {et~al.}(2004)\citenamefont
  {Saltelli}, \citenamefont {Tarantola}, \citenamefont {Campolongo},\ and\
  \citenamefont {Ratto}}]{Saltelli2004}%
  \BibitemOpen
  \bibfield  {author} {\bibinfo {author} {\bibfnamefont {A.}~\bibnamefont
  {Saltelli}}, \bibinfo {author} {\bibfnamefont {S.}~\bibnamefont {Tarantola}},
  \bibinfo {author} {\bibfnamefont {F.}~\bibnamefont {Campolongo}}, \ and\
  \bibinfo {author} {\bibfnamefont {M.}~\bibnamefont {Ratto}},\ }\href@noop {}
  {\emph {\bibinfo {title} {{Sensitivity analysis in practice}}}}\ (\bibinfo
  {publisher} {{John Wiley \& Sons Ltd.}},\ \bibinfo {year} {2004})\BibitemShut
  {NoStop}%
\bibitem [{\citenamefont {Erguler}\ and\ \citenamefont
  {Stumpf}(2011)}]{Erguler2011}%
  \BibitemOpen
  \bibfield  {author} {\bibinfo {author} {\bibfnamefont {K.~K.}\ \bibnamefont
  {Erguler}}\ and\ \bibinfo {author} {\bibfnamefont {M.~P.~H.}\ \bibnamefont
  {Stumpf}},\ }\bibfield  {title} {\enquote {\bibinfo {title} {{Practical
  limits for reverse engineering of dynamical systems: a statistical analysis
  of sensitivity and parameter inferability in systems biology models.}}}\
  }\href@noop {} {\bibfield  {journal} {\bibinfo  {journal} {Molecular
  bioSystems}\ }\textbf {\bibinfo {volume} {7}},\ \bibinfo {pages} {1593--1602}
  (\bibinfo {year} {2011})}\BibitemShut {NoStop}%
\bibitem [{\citenamefont {Gunawan}\ \emph {et~al.}(2005)\citenamefont
  {Gunawan}, \citenamefont {Cao}, \citenamefont {Petzold},\ and\ \citenamefont
  {Doyle}}]{Gunawan2005}%
  \BibitemOpen
  \bibfield  {author} {\bibinfo {author} {\bibfnamefont {R.}~\bibnamefont
  {Gunawan}}, \bibinfo {author} {\bibfnamefont {Y.}~\bibnamefont {Cao}},
  \bibinfo {author} {\bibfnamefont {L.}~\bibnamefont {Petzold}}, \ and\
  \bibinfo {author} {\bibfnamefont {F.~J.}\ \bibnamefont {Doyle}},\ }\bibfield
  {title} {\enquote {\bibinfo {title} {{Sensitivity analysis of discrete
  stochastic systems.}}}\ }\href@noop {} {\bibfield  {journal} {\bibinfo
  {journal} {Biophysical Journal}\ }\textbf {\bibinfo {volume} {88}},\ \bibinfo
  {pages} {2530--2540} (\bibinfo {year} {2005})}\BibitemShut {NoStop}%
\bibitem [{\citenamefont {Plyasunov}\ and\ \citenamefont
  {Arkin}(2007)}]{Plyasunov2007}%
  \BibitemOpen
  \bibfield  {author} {\bibinfo {author} {\bibfnamefont {S.}~\bibnamefont
  {Plyasunov}}\ and\ \bibinfo {author} {\bibfnamefont {A.~P.}\ \bibnamefont
  {Arkin}},\ }\bibfield  {title} {\enquote {\bibinfo {title} {{Efficient
  stochastic sensitivity analysis of discrete event systems}},}\ }\href@noop {}
  {\bibfield  {journal} {\bibinfo  {journal} {Journal of Computational
  Physics}\ }\textbf {\bibinfo {volume} {221}},\ \bibinfo {pages} {724--738}
  (\bibinfo {year} {2007})}\BibitemShut {NoStop}%
\bibitem [{\citenamefont {Komorowski}, \citenamefont {Zurauskiene},\ and\
  \citenamefont {Stumpf}(2012)}]{Komorowski2012b}%
  \BibitemOpen
  \bibfield  {author} {\bibinfo {author} {\bibfnamefont {M.}~\bibnamefont
  {Komorowski}}, \bibinfo {author} {\bibfnamefont {J.}~\bibnamefont
  {Zurauskiene}}, \ and\ \bibinfo {author} {\bibfnamefont {M.~P.~H.}\
  \bibnamefont {Stumpf}},\ }\bibfield  {title} {\enquote {\bibinfo {title}
  {{StochSens--matlab package for sensitivity analysis of stochastic chemical
  systems.}}}\ }\href@noop {} {\bibfield  {journal} {\bibinfo  {journal}
  {Bioinformatics (Oxford, England)}\ }\textbf {\bibinfo {volume} {28}},\
  \bibinfo {pages} {731--733} (\bibinfo {year} {2012})}\BibitemShut {NoStop}%
\bibitem [{\citenamefont {Sheppard}, \citenamefont {Rathinam},\ and\
  \citenamefont {Khammash}(2012)}]{Sheppard2012}%
  \BibitemOpen
  \bibfield  {author} {\bibinfo {author} {\bibfnamefont {P.~W.}\ \bibnamefont
  {Sheppard}}, \bibinfo {author} {\bibfnamefont {M.}~\bibnamefont {Rathinam}},
  \ and\ \bibinfo {author} {\bibfnamefont {M.}~\bibnamefont {Khammash}},\
  }\bibfield  {title} {\enquote {\bibinfo {title} {{SPSens: A software package
  for stochastic parameter sensitivity analysis of biochemical reaction
  networks.}}}\ }\href@noop {} {\bibfield  {journal} {\bibinfo  {journal}
  {Bioinformatics (Oxford, England)}\ ,\ \bibinfo {pages} {--}} (\bibinfo
  {year} {2012})}\BibitemShut {NoStop}%
\bibitem [{\citenamefont {Varma}, \citenamefont {Morbidelli},\ and\
  \citenamefont {Wu}(1999)}]{Varma1999}%
  \BibitemOpen
  \bibfield  {author} {\bibinfo {author} {\bibfnamefont {A.}~\bibnamefont
  {Varma}}, \bibinfo {author} {\bibfnamefont {M.}~\bibnamefont {Morbidelli}}, \
  and\ \bibinfo {author} {\bibfnamefont {H.~H.}\ \bibnamefont {Wu}},\
  }\href@noop {} {\emph {\bibinfo {title} {Parametric sensitivity in chemical
  systems}}}\ (\bibinfo  {publisher} {Cambridge, U.K. ; New York, NY :
  Cambridge University Press},\ \bibinfo {year} {1999})\BibitemShut {NoStop}%
\bibitem [{\citenamefont {Serban}\ and\ \citenamefont
  {Hindmarsh}(2003)}]{Serban2003}%
  \BibitemOpen
  \bibfield  {author} {\bibinfo {author} {\bibfnamefont {R.}~\bibnamefont
  {Serban}}\ and\ \bibinfo {author} {\bibfnamefont {A.~C.}\ \bibnamefont
  {Hindmarsh}},\ }\bibfield  {title} {\enquote {\bibinfo {title} {{CVODES: An
  ODE solver with sensitivity analysis capabilities}},}\ }in\ \href@noop {}
  {\emph {\bibinfo {booktitle} {Preprint UCRL-JP-200039, Lawrence Livermore
  National Laboratory}}}\ (\bibinfo {year} {2003})\BibitemShut {NoStop}%
\bibitem [{\citenamefont {Gelman}\ \emph {et~al.}(2003)\citenamefont {Gelman},
  \citenamefont {J.B.}, \citenamefont {Stern},\ and\ \citenamefont
  {Rubin}}]{Gelman:2003}%
  \BibitemOpen
  \bibfield  {author} {\bibinfo {author} {\bibfnamefont {A.}~\bibnamefont
  {Gelman}}, \bibinfo {author} {\bibfnamefont {C.}~\bibnamefont {J.B.}},
  \bibinfo {author} {\bibfnamefont {H.}~\bibnamefont {Stern}}, \ and\ \bibinfo
  {author} {\bibfnamefont {D.}~\bibnamefont {Rubin}},\ }\href@noop {} {\emph
  {\bibinfo {title} {Bayesian Data Analysis}}},\ \bibinfo {edition} {2nd}\ ed.\
  (\bibinfo  {publisher} {Chapman \& Hall/CRC},\ \bibinfo {year}
  {2003})\BibitemShut {NoStop}%
\bibitem [{\citenamefont {Toni}\ \emph {et~al.}(2009)\citenamefont {Toni},
  \citenamefont {Welch}, \citenamefont {Strelkowa}, \citenamefont {Ipsen},\
  and\ \citenamefont {Stumpf}}]{Toni:2008aa}%
  \BibitemOpen
  \bibfield  {author} {\bibinfo {author} {\bibfnamefont {T.}~\bibnamefont
  {Toni}}, \bibinfo {author} {\bibfnamefont {D.}~\bibnamefont {Welch}},
  \bibinfo {author} {\bibfnamefont {N.}~\bibnamefont {Strelkowa}}, \bibinfo
  {author} {\bibfnamefont {D.}~\bibnamefont {Ipsen}}, \ and\ \bibinfo {author}
  {\bibfnamefont {M.}~\bibnamefont {Stumpf}},\ }\bibfield  {title} {\enquote
  {\bibinfo {title} {Approximate bayesian computation scheme for parameter
  inference and model selection in dynamical systems},}\ }\href@noop {}
  {\bibfield  {journal} {\bibinfo  {journal} {J.Roy.Soc. Interface}\ }\textbf
  {\bibinfo {volume} {6}},\ \bibinfo {pages} {187--202} (\bibinfo {year}
  {2009})}\BibitemShut {NoStop}%
\bibitem [{\citenamefont {{Chevalier}}\ and\ \citenamefont
  {{El-Samad}}(2011)}]{Chevalier2011}%
  \BibitemOpen
  \bibfield  {author} {\bibinfo {author} {\bibfnamefont {M.~W.}\ \bibnamefont
  {{Chevalier}}}\ and\ \bibinfo {author} {\bibfnamefont {H.}~\bibnamefont
  {{El-Samad}}},\ }\bibfield  {title} {\enquote {\bibinfo {title} {A
  data-integrated method for analyzing stochastic biochemical networks},}\
  }\href@noop {} {\bibfield  {journal} {\bibinfo  {journal} {The Journal of
  Chemical Physics}\ ,\ \bibinfo {pages} {214110}} (\bibinfo {year}
  {2011})}\BibitemShut {NoStop}%
\bibitem [{\citenamefont {{Azunre}}, \citenamefont {{G\`omez-Uribe}},\ and\
  \citenamefont {{Verghese}}(2011)}]{Azunre2011}%
  \BibitemOpen
  \bibfield  {author} {\bibinfo {author} {\bibfnamefont {P.}~\bibnamefont
  {{Azunre}}}, \bibinfo {author} {\bibfnamefont {C.}~\bibnamefont
  {{G\`omez-Uribe}}}, \ and\ \bibinfo {author} {\bibfnamefont {G.}~\bibnamefont
  {{Verghese}}},\ }\bibfield  {title} {\enquote {\bibinfo {title} {Mass
  fluctuation kinetics: analysis and computation of equilibria and local
  dynamics},}\ }\href@noop {} {\bibfield  {journal} {\bibinfo  {journal} {IET
  Systems Biology}\ }\textbf {\bibinfo {volume} {5}},\ \bibinfo {pages}
  {325--335} (\bibinfo {year} {2011})}\BibitemShut {NoStop}%
\bibitem [{\citenamefont {{Milner}}, \citenamefont {{Gilespie}},\ and\
  \citenamefont {{Wilkinson}}(2012)}]{Milner2011b}%
  \BibitemOpen
  \bibfield  {author} {\bibinfo {author} {\bibfnamefont {P.}~\bibnamefont
  {{Milner}}}, \bibinfo {author} {\bibfnamefont {C.~C.}\ \bibnamefont
  {{Gilespie}}}, \ and\ \bibinfo {author} {\bibfnamefont {D.~J.}\ \bibnamefont
  {{Wilkinson}}},\ }\bibfield  {title} {\enquote {\bibinfo {title} {Moment
  closure based parameter inference of stochastic kinetic models},}\
  }\href@noop {} {\bibfield  {journal} {\bibinfo  {journal} {Statistics and
  Computing}\ }\textbf {\bibinfo {volume} {231}},\ \bibinfo {pages} {99--104}
  (\bibinfo {year} {2012})}\BibitemShut {NoStop}%
\bibitem [{\citenamefont {{Kuegler}}(2012)}]{Kuegler2012}%
  \BibitemOpen
  \bibfield  {author} {\bibinfo {author} {\bibfnamefont {P.}~\bibnamefont
  {{Kuegler}}},\ }\bibfield  {title} {\enquote {\bibinfo {title} {Moment
  fitting for parameter inference in repeatedly and partially observed
  stochastic biological models},}\ }\href@noop {} {\bibfield  {journal}
  {\bibinfo  {journal} {Plos one}\ }\textbf {\bibinfo {volume} {7}},\ \bibinfo
  {pages} {1--15} (\bibinfo {year} {2012})}\BibitemShut {NoStop}%
\bibitem [{\citenamefont {Toni}\ \emph {et~al.}(2012)\citenamefont {Toni},
  \citenamefont {Ozaki}, \citenamefont {Kirk}, \citenamefont {Kuroda},\ and\
  \citenamefont {Stumpf}}]{Toni:2012bo}%
  \BibitemOpen
  \bibfield  {author} {\bibinfo {author} {\bibfnamefont {T.}~\bibnamefont
  {Toni}}, \bibinfo {author} {\bibfnamefont {Y.-i.}\ \bibnamefont {Ozaki}},
  \bibinfo {author} {\bibfnamefont {P.}~\bibnamefont {Kirk}}, \bibinfo {author}
  {\bibfnamefont {S.}~\bibnamefont {Kuroda}}, \ and\ \bibinfo {author}
  {\bibfnamefont {M.~P.~H.}\ \bibnamefont {Stumpf}},\ }\bibfield  {title}
  {\enquote {\bibinfo {title} {{Elucidating the in vivo phosphorylation
  dynamics of the ERK MAP kinase using quantitative proteomics data and
  Bayesian model selection.}}}\ }\href@noop {} {\bibfield  {journal} {\bibinfo
  {journal} {Molecular BioSystems}\ }\textbf {\bibinfo {volume} {8}},\ \bibinfo
  {pages} {1921--1929} (\bibinfo {year} {2012})}\BibitemShut {NoStop}%
\bibitem [{\citenamefont {{Zechner}}\ \emph {et~al.}(2012)\citenamefont
  {{Zechner}}, \citenamefont {{Ruess}}, \citenamefont {{Krenn}}, \citenamefont
  {{Pelet}}, \citenamefont {{Peter}}, \citenamefont {{Lygeros}},\ and\
  \citenamefont {{Koeppl}}}]{Zechner2012}%
  \BibitemOpen
  \bibfield  {author} {\bibinfo {author} {\bibfnamefont {C.}~\bibnamefont
  {{Zechner}}}, \bibinfo {author} {\bibfnamefont {J.}~\bibnamefont {{Ruess}}},
  \bibinfo {author} {\bibfnamefont {P.}~\bibnamefont {{Krenn}}}, \bibinfo
  {author} {\bibfnamefont {S.}~\bibnamefont {{Pelet}}}, \bibinfo {author}
  {\bibfnamefont {M.}~\bibnamefont {{Peter}}}, \bibinfo {author} {\bibfnamefont
  {J.}~\bibnamefont {{Lygeros}}}, \ and\ \bibinfo {author} {\bibfnamefont
  {H.}~\bibnamefont {{Koeppl}}},\ }\bibfield  {title} {\enquote {\bibinfo
  {title} {Moment-based inference predicts bimodality in transient gene
  expression},}\ }\href@noop {} {\bibfield  {journal} {\bibinfo  {journal}
  {PNAS}\ } (\bibinfo {year} {2012})}\BibitemShut {NoStop}%
\bibitem [{\citenamefont {{Salehi-Reyhani}}\ \emph {et~al.}(2011)\citenamefont
  {{Salehi-Reyhani}}, \citenamefont {{Kaplinsky}}, \citenamefont {{Burgin}},
  \citenamefont {{Novakova}}, \citenamefont {Andrew J.~{deMello}},
  \citenamefont {{Parker}}, \citenamefont {{Neil}}, \citenamefont {{Ces}},
  \citenamefont {{French}}, \citenamefont {R.{Willison}},\ and\ \citenamefont
  {{Klug}}}]{Klug2011}%
  \BibitemOpen
  \bibfield  {author} {\bibinfo {author} {\bibfnamefont {A.}~\bibnamefont
  {{Salehi-Reyhani}}}, \bibinfo {author} {\bibfnamefont {J.}~\bibnamefont
  {{Kaplinsky}}}, \bibinfo {author} {\bibfnamefont {E.}~\bibnamefont
  {{Burgin}}}, \bibinfo {author} {\bibfnamefont {M.}~\bibnamefont
  {{Novakova}}}, \bibinfo {author} {\bibfnamefont {R.~H.~T.}\ \bibnamefont
  {Andrew J.~{deMello}}}, \bibinfo {author} {\bibfnamefont {P.}~\bibnamefont
  {{Parker}}}, \bibinfo {author} {\bibfnamefont {M.~A.~A.}\ \bibnamefont
  {{Neil}}}, \bibinfo {author} {\bibfnamefont {O.}~\bibnamefont {{Ces}}},
  \bibinfo {author} {\bibfnamefont {P.}~\bibnamefont {{French}}}, \bibinfo
  {author} {\bibfnamefont {K.}~\bibnamefont {R.{Willison}}}, \ and\ \bibinfo
  {author} {\bibfnamefont {D.}~\bibnamefont {{Klug}}},\ }\bibfield  {title}
  {\enquote {\bibinfo {title} {A first step towards practical single cell
  proteomics: a microfluidic antibody capture chip with tirf detection},}\
  }\href@noop {} {\bibfield  {journal} {\bibinfo  {journal} {Lab on a Chip}\ ,\
  \bibinfo {pages} {1256--1261}} (\bibinfo {year} {2011})}\BibitemShut
  {NoStop}%
\bibitem [{\citenamefont {{Lin}}\ \emph {et~al.}(2011)\citenamefont {{Lin}},
  \citenamefont {{Trouillon}}, \citenamefont {{Safina}},\ and\ \citenamefont
  {{Ewing}}}]{Lin2011}%
  \BibitemOpen
  \bibfield  {author} {\bibinfo {author} {\bibfnamefont {Y.}~\bibnamefont
  {{Lin}}}, \bibinfo {author} {\bibfnamefont {R.}~\bibnamefont {{Trouillon}}},
  \bibinfo {author} {\bibfnamefont {G.}~\bibnamefont {{Safina}}}, \ and\
  \bibinfo {author} {\bibfnamefont {A.~G.}\ \bibnamefont {{Ewing}}},\
  }\bibfield  {title} {\enquote {\bibinfo {title} {Chemical analysis of single
  cells},}\ }\href@noop {} {\bibfield  {journal} {\bibinfo  {journal}
  {Analytical Chemistry}\ }\textbf {\bibinfo {volume} {83}},\ \bibinfo {pages}
  {4369Ð4392} (\bibinfo {year} {2011})}\BibitemShut {NoStop}%
\bibitem [{\citenamefont {Ozaki}\ \emph {et~al.}(2010)\citenamefont {Ozaki},
  \citenamefont {Uda}, \citenamefont {Saito}, \citenamefont {Chung},
  \citenamefont {Kubota},\ and\ \citenamefont {Kuroda}}]{Ozaki:2010p26116}%
  \BibitemOpen
  \bibfield  {author} {\bibinfo {author} {\bibfnamefont {Y.-i.}\ \bibnamefont
  {Ozaki}}, \bibinfo {author} {\bibfnamefont {S.}~\bibnamefont {Uda}}, \bibinfo
  {author} {\bibfnamefont {T.~H.}\ \bibnamefont {Saito}}, \bibinfo {author}
  {\bibfnamefont {J.}~\bibnamefont {Chung}}, \bibinfo {author} {\bibfnamefont
  {H.}~\bibnamefont {Kubota}}, \ and\ \bibinfo {author} {\bibfnamefont
  {S.}~\bibnamefont {Kuroda}},\ }\bibfield  {title} {\enquote {\bibinfo {title}
  {{A quantitative image cytometry technique for time series or population
  analyses of signaling networks}},}\ }\href@noop {} {\bibfield  {journal}
  {\bibinfo  {journal} {PLoS One}\ }\textbf {\bibinfo {volume} {5}},\ \bibinfo
  {pages} {e9955} (\bibinfo {year} {2010})}\BibitemShut {NoStop}%
\end{thebibliography}
\end{document}